\documentclass[aps,prd,twocolumn,groupedaddress]{revtex4-1}
\usepackage{graphicx}
\include{amssym}

\begin{document}
\title{Hadronic electroweak current and $\pi - a_1$ mixing}

\author{A. A. Osipov$^{1,2}$}
\email[]{aaosipov@jinr.ru}

\author{B. Hiller$^{3}$}
\email[]{brigitte@fis.uc.pt}

\author{P. M. Zhang$^{1,4}$}
\email[]{zhpm@impcas.ac.cn}

\affiliation{$^1$Institute of Modern Physics, Chinese Academy of Sciences, Lanzhou 730000, China}

\affiliation{$^2$Bogoliubov Laboratory of Theoretical Physics, Joint Institute for Nuclear Research, Dubna, 141980, Russia}

\affiliation{$^3$CFisUC, Department of Physics, University of Coimbra, P-3004-516 Coimbra, Portugal}

\affiliation{$^4$University of Chinese Academy of Sciences, Yuquanlu 19A, Beijing 100049, China}

\begin{abstract}
It is found that in presence of electroweak interactions the gauge covariant diagonalization of the axial-vector -- pseudoscalar mixing in the effective meson Lagrangian leads to a deviation from the vector meson and the axial-vector meson dominance of the entire hadronic electroweak current. The essential features of such a modification of the theory are investigated in the framework of the extended Nambu-Jona-Lasinio model with explicit breaking of chiral $U(2) \times U(2)$ symmetry. The Schwinger-DeWitt method is used as a major tool in our study of the real part of the relevant effective action. Some straightforward applications are considered.
\end{abstract}

\maketitle

\section{Introduction}
At low energies, the strong and electroweak interactions of mesons can be described by effective chiral Lagrangians \cite{Weinberg67,Gasiorovicz69,Weinberg79,Gasser83,Volkov84,Gasser84,Gasser85,Volkov86,Ebert86,Meissner88,Bando88,Ecker89plb,Ecker89npb}. The hadronic part of these Lagrangians can be constructed explicitly in terms of meson variables, or by using microscopic theories of quarks with subsequent bosonization. The Nambu-Jona-Lasinio (NJL) model \cite{Nambu61a,Nambu61b}, reinterpreted in terms of the quark fields \cite{Ebert82,Ebert83,Volkov84}, is one of such microscopic approaches. The effective chiral Lagrangians are known as a means for the treatment of the approximate chiral symmetry of QCD in the study of particle physics. In particular, in \cite{Ebert82,Ebert83,Ebert86} chiral symmetry was found to confirm the phenomenologically successful ideas of vector meson dominance (VMD) \cite{Nambu57,Frazer60,Gell-Mann61,Gell-Mann62,Nambu62,Kroll67,Sakurai69}, axial-vector dominance of the charged hadronic weak current, and vector/axial-vector universality.

The idea of the existence of the isoscalar vector $\omega$-meson was suggested by Nambu to explain the behaviour of the isoscalar part of the electromagnetic nucleon form factor \cite{Nambu57}. A bit later, Frazer and Fulco have predicted the isovector vector states, the $\rho$-mesons, and have shown that the $\gamma\to\rho^0 \to N\bar N$ transition can describe the isovector part of the nucleon form factor. These discoveries are the cradle of the vector dominance idea. Soon after that \cite{Gell-Mann61,Gell-Mann62,Nambu62}  the idea has been elaborated to apply to all electromagnetic interactions of hadrons. An explicit statement of the meaning of vector dominance in the language of a local Lagrangian field theory has been suggested by Kroll, Lee, and Zumino in \cite{Kroll67} in the following form: "To a very good approximation the entire hadronic electromagnetic current operator is identical with a linear combination of the known neutral vector-meson fields". For the theories originally formulated in terms of quark fields it means that vector fields can be introduced in such a way that direct interactions of photons with quarks vanish. As a result, the theory does not have direct couplings of mesons with photons except for the $\omega, \rho^0, \phi \to \gamma$ transitions. A similar statement can be formulated in relation with the charged hadronic weak current, where the charged axial-vector mesons may play a dominant role in the description of the weak interactions of hadrons. It has been also realized \cite{Sakurai69} that universality is a direct consequence of the complete dominance of the vector $\rho^0$ meson in the isovector form factor.         

Recently attention was drawn to the need of gauge covariant pseudoscalar - axial-vector ($p -a_\mu$) diagonalization of the effective meson Lagrangian in presence of electromagnetic interactions \cite{Osipov18jl,Osipov18vd}. The standard replacement
\begin{equation}
\label{noncd}
a_{\mu}\to a_{\mu}+ \kappa m \partial_\mu p
\end{equation}
[where $\kappa$ is a constant, $m$ is the constituent quark mass, $a_\mu =a_{\mu a}\tau_a, \,\, p=p_a\tau_a $ with $a=0,1,2,3$,  $\tau_0=1$, and $\tau_i$ $(i=1,2,3)$ are the Pauli matrices] ruins the $U(1)$ gauge covariant transformation properties of the axial-vector field
\begin{equation}
a_\mu\to a_\mu' =e^{ie\alpha Q}a_\mu e^{-ie\alpha Q},
\end{equation}
where $\alpha (x)$ is a local phase, and gives rise to all sorts of gauge symmetry breaking. As an example, it has been shown that the amplitude of the anomalous decay $a_1(1260)\to\gamma\pi^+\pi^-$ is not gauge invariant if the standard approach is used. It has been also noticed that a covariant diagonalization
\begin{equation}
\label{covd}
a_{\mu}\to a_{\mu}+ \kappa m {\cal D}_\mu p,
\end{equation}
where ${\cal D}_\mu p=\partial_\mu p -ie{\cal A}_\mu [Q,p]$,  and ${\cal A}_\mu$ is the electromagnetic field, leads to a gauge invariant result, but the price is a deviation from the VMD picture [the covariant derivative induces a coupling of the quark-antiquark pair with the charged pseudoscalar and the photon fields, which is the source of the new non VMD type electromagnetic interactions]. This result has physical consequences, and therefore should be very carefully scrutinized.

In this context, let us also recall the old result of the paper \cite{Kugo85}, based on the general solutions of the Wess-Zumino anomaly equation which incorporate vector mesons as dynamical gauge bosons of the hidden local symmetry in the nonlinear chiral Lagrangian. The authors have found that the "complete vector meson dominance" hypotesis of photon couplings is invalid in either $\pi^0\to\gamma \gamma$ or $\gamma\to\pi\pi\pi$ processes. The covariant diagonalization provides us with an additional reasoning according to which the VMD is not the whole story for electromagnetic interactions of mesons.

In the present paper we extend the idea of covariant diagonalization to the case of electroweak interactions governed by the non-abelian group of local $SU(2)_L\times U(1)_R$ symmetry. For that, as in \cite{Osipov18jl,Osipov18vd}, we use the NJL Lagrangian with the global $U(2)_L\times U(2)_R$ chiral symmetric spin zero and spin one four-quark forces.  Such extension is not so straightforward as one might expect from the pure electromagnetic case. Indeed, one cannot modify Eq. (\ref{covd}) by simply replacing the $U(1)$ covariant derivative ${\cal D}_\mu p$ by the corresponding $SU(2)_L\times U(1)_R$ gauge covariant derivative of the pseudoscalar field. 

First, one should find the $U(2)_L\times U(2)_R$ chiral covariant analog for the derivative $\partial_\mu p$, and only after that replace the usual derivatives by the gauge covariant ones. Fortunately, this step is known \cite{Osipov00,Osipov02}, but it requires the following modifications both of the axial-vector and the vector fields. In the spontaneously broken phase the replacement has the form
\begin{eqnarray}
\label{nongcov}
a_\mu&\to& a_\mu +\frac{\kappa}{2} \left(\{p,\partial_\mu \bar s\} -\{\bar s,\partial_\mu p \} \right)  \\
v_\mu&\to& v_\mu +i\,\frac{\kappa}{2}\left([p,\partial_\mu p] + [\bar s, \partial_\mu \bar s] \right),
\end{eqnarray}
where the scalar field $s$ in the Nambu-Goldstone vacuum is given by $\bar s =s-m$. Notice that the two bilinear combinations of scalar and pseudoscalar fields transform like axial-vector and vector fields and are chiral partners with respect to the linear transformations of the chiral group. 

As a second step, one should show that these bilinear combinations, after the replacement of usual derivatives of meson fields by the gauge covariant ones, possess the gauge covariant transformation laws that preserve the $SU(2)_L\times U(1)_R$ gauge symmetry of the Lagrangian. Only then can one be sure that the gauge symmetry of the electroweak interactions is properly incorporated in the effective meson Lagrangian. 

All these steps are considered in the present paper in detail. As a result, we obtain the effective Lagrangian which describes the electroweak interactions of mesons in accord with the gauge symmetry requirements. The various phenomenological applications of this Lagrangian will be considered elsewhere.

The purpose of this paper is to exhibit a Lagrangian theory in which the effective electroweak interactions of mesons may deviate from the phenomenologically successful ideas of VMD and axial-vector dominance due to the $p-a_\mu$ mixing and gauge symmetry requirements. In Sec. \ref{L} we recall the construction of the $SU(2)_L\times U(1)_R$ gauge covariant derivative of the quark fields and describe the version of the NJL model with $U(2)_L\times U(2)_R$ chiral symmetric interactions which is used in our studies. In Sec \ref{gtmf} we construct the gauge transformations of meson fields and compare them with their chiral transformations. In Sec. \ref{gcd} we find the gauge covariant derivatives of pseudoscalar and scalar fields. In Sec. \ref{cd} we show that the gauge covariant replacement of vector and axial vector fields violates neither gauge nor chiral symmetries, solving the problem of $\pi -a_1$ diagonalization of the chiral Lagrangian. In Sec. \ref{vavd} we show that covariant diagonalization adds to the standard picture of vector (axial-vector) dominance a set of new interactions while keeping the old results unchanged. Here we also obtain the neutral hadronic weak current, which has not been considered in \cite{Ebert86}. In Sec. \ref{sdwe} we bosonize the theory by using the Schwinger-DeWitt method to derive the one-quark-loop contribution to the real part of the effective action in the long-wavelength approximation. The one photon interactions following from the total effective Lagrangian are considered in Sec. \ref{ei}. Here we present three useful examples which show the common features and differences between covariant and non-covariant evaluation of the electromagnetic vertices $\gamma\pi\pi$, $a_1\gamma\pi$, and the $\rho^0\to \gamma\pi^+\pi^-$ decay amplitude. In Sec. \ref{Zbi} the one $Z$ boson part of the Lagrangian is considered. Here we show some parity-violating processes with the isospin change $\Delta I=1$. The hadronic charged weak current is shown in Sec. \ref{Wbi}. In Sec. \ref{concl} we summarize our results and discuss the possible role of the non VMD (and axial-vector dominance) electroweak contributions for the theory.

\section{Electroweak-Quark Lagrangian}
\label{L}
Our starting point is the following quark NJL Lagrangian density with the global $U(2)_R\times U(2)_L$ chiral symmetric four-quark interactions, which also possesses the gauge $SU(2)_L\times U(1)_R$ symmetry of the electroweak interactions
\begin{eqnarray}
\label{lag}
&&{\cal L}=\bar q(i\gamma^\mu{\cal D}_\mu -\hat m)q + {\cal L}_{S} + {\cal L}_{V}+{\cal L}_{EW}, \\
\label{lagsp}
&&{\cal L}_{S}=(G_S/2)\left[(\bar q\tau_a q)^2+(\bar qi\gamma_5\tau_a q)^2 \right], \\
\label{lagva}
&&{\cal L}_{V}=-(G_V/2)\left[(\bar q\gamma^\mu\tau_a q)^2+(\bar q\gamma^\mu\gamma_5\tau_a q)^2 \right], \\
\label{lagem}
&&{\cal L}_{EW}={\cal L}_{\mbox{\footnotesize gauge}}+{\cal L}_{\mbox{\footnotesize Higgs}}+{\cal L}_{\mbox{\footnotesize FP}}+\mbox{gauge fixing}.
\end{eqnarray}
Here, in the notation of the quark fields $q(x)$ the color and flavor indices are suppressed. The quark part of the Lagrangian density includes both spin-0 and spin-1 four-quark couplings with dimensional constants $G_S$ and $G_V$ correspondingly; the matrix $\hat m$ has equal diagonal elements $\hat m_u=\hat m_d$ corresponding to the current masses of $u$ and $d$ quarks in the isospin limit; $\tau_a=(\tau_0,\vec\tau)$ for $a=0,1,2,3$, where $\tau_0$ is a $2\times 2$ unit matrix, and $\vec\tau$ are the $SU(2)$ Pauli matrices; $\gamma^\mu$ are the standard Dirac matrices in four dimensions. The covariant [with respect to the action of the gauge group of the electroweak interactions] derivative of quarks is given by
\begin{equation}
{\cal D}_\mu q = \left[\partial_\mu -igA_\mu P_L -ig'B_\mu (Q-T_3P_L) \right]q,
\end{equation}
where the matrix $Q=T_3+Y_L=1/2(\tau_3 +1/3)$ accumulates the electromagnetic charges of $u$ and $d$ quarks in relative units of the proton charge $e>0$; $A_\mu=A_\mu^i T_i$ and $B_\mu$ are gauge fields of the $SU(2)_L$ and $U(1)_R$ groups of local transformations, correspondingly; $T_i=\tau_i/2$; $P_{L,R}=(1\mp\gamma_5)/2$.

The physical variables $Z_\mu, {\cal A}_\mu$ of the neutral gauge fields can be introduced through the orthogonal transformation
\begin{eqnarray}
&& Z_\mu =  \cos \theta_W A_\mu^3 -\sin\theta_W B_\mu ,  \nonumber  \\
&& {\cal A}_\mu =  \sin\theta_W A_\mu^3 + \cos \theta_W B_\mu ,
\end{eqnarray}
where the Weinberg angle $\theta_W$ is defined by the couplings of electroweak interactions $\cos \theta_W = g/\sqrt{g^2+(g')^2}$, and $\sin \theta_W = g'/\sqrt{g^2+(g')^2}$. The charged physical states of gauge fields are $W_\mu^\pm=\left(A_\mu^1\mp iA_\mu^2\right)/\sqrt{2}$.

The covariant derivative of the quark field can be rewritten in terms of the physical states as follows
\begin{eqnarray}
&&{\cal D_\mu}q=\left[\partial_\mu -ieQ {\cal A}_\mu - \frac{igZ_\mu}{\cos\theta_W}\left(T_3P_L-\sin^2\theta_WQ\right) \right. \nonumber \\
&&  \qquad\qquad\quad  -\left. ig P_L\left(T_+W^+_\mu +T_- W^-_\mu \right)\right]q,
\end{eqnarray}
where $T_\pm = (T_1\pm iT_2)/\sqrt{2}$, and $\sin^2\theta_W=0.23$.

In the following, we will not use  the Lagrangian density ${\cal L}_{EW}$. For this reason its expression is not given. The term "gauge fixing" there contains the Faddeev-Popov ghosts and the corresponding gauge fixing terms.

The Lagrangian density ${\cal L}$ is of the symmetry breaking type, in the sense that starting from some critical value of $G_S$ the minimum of the effective potential occurs for non-zero values of $\langle \bar q q\rangle\neq 0$ and the constituent quark mass $m$. This is the chiral symmetry breaking phenomenon. In the non-symmetric vacuum, the physical spectrum contains $q\bar q$ bound states. Therefore, it is convenient to introduce meson variables in the corresponding functional integral explicitly. This can be done by transforming the nonlinear interactions of quarks to the Yukawa interactions of quarks with auxiliary boson fields
\begin{eqnarray}
\label{smat3}
S [{\cal A}_\mu, Z_\mu, &&W_\mu^\pm ]=\int [dq] [d\bar q] [ds_a] [dp_a] [dv_{a\mu} ]
   [d{a}_{a\mu} ]  \nonumber \\
&&\times\exp i\!\!\int\!\! d^4x\left(\bar q\, D_mq +\mathcal L_M +{\cal L}_{EW}\right).
\end{eqnarray}
Here $D_m$ is the Dirac operator in presence of background fields
\begin{equation}
D_m = i\gamma^\mu\mathcal D_\mu -m+s+i\gamma_5 p +\gamma^\mu v_\mu
+ \gamma^\mu\gamma_5 a_\mu .
\end{equation}
The scalar, pseudoscalar, vector and axial vector fields are $s=s_a\tau_a,\, p=p_a\tau_a,\, v_{\mu } =v_{a\mu }\tau_a,\, a_\mu =a_{a\mu }\tau_a$. $\mathcal L_M$ describes the meson mass part of the Lagrangian density
\begin{eqnarray}
\label{Mpart1}
\mathcal L_M =&-&\frac{1}{4G_S}\mbox{tr} \left[ (s -m+\hat m)^2+p^{2}\right] \nonumber \\
                        &+&\frac{1}{4G_V} \mbox{tr} \left( v_\mu^{2}+a_\mu^{2}\right),
\end{eqnarray}
where the trace is taken over flavor indices.

The spontaneous symmetry breakdown leads to the mixing of the pseudoscalar, $p$, and axial-vector, $a_\mu$, fields through the one quark loop. The mixing term occurs to be proportional to $\mbox{tr}(a_\mu\partial_\mu p)$. To avoid mixing one usually redefines the axial-vector field by the replacement (\ref{noncd}), where the coupling $\kappa$ is unambiguously fixed to avoid mixing. This conventional method has the advantage of putting the Lagrangian in a useful form in the most economic and practical way possible. However, this does not work when the Lagrangian possesses both the chiral and the gauge symmetries. In this case the replacement must not violate either symmetries, and this imposes a very powerful restriction on the form of the second term in (\ref{noncd}).

We now show how to construct a chiral and gauge covariant replacement out of the spin-0 and gauge fields.

\section{Gauge Transformations of Meson Fields}
\label{gtmf}

The first step toward the solution is to find the gauge transformation law of meson fields. The $SU(2)_L\times U(1)_R$ gauge group acts on the quark fields by the following infinitesimal transformations
\begin{eqnarray}
\delta q_R &=& ie\alpha Qq_R, \nonumber \\
\delta\bar q_R&=&-ie\alpha \bar q_R Q, \nonumber \\
\delta q_L &=& i(\omega +e\alpha Y_L) q_L,  \nonumber \\
\delta \bar q_L&=& -i\bar q_L(\omega + e\alpha Y_L),
\end{eqnarray}
where $q_R=P_R q$, $q_L=P_L q$, and  $\alpha$ and $\omega =\omega_iT_i$ are the local parameters of the $U(1)_R$ and $SU(2)_L$ gauge transformations correspondingly.

The gauge fields can easily be seen to transform under $SU(2)_L\times U(1)_R$ according to
\begin{eqnarray}
&& \delta B_\mu = \frac{e}{g'}\,\partial_\mu \alpha , \\
&& \delta A_\mu = i[\omega, A_\mu ] + \frac{1}{g} \,\partial_\mu \omega.
\end{eqnarray}

One can show that the Lagrangian densities ${\cal L}_S$ and ${\cal L}_V$ are invariant with respect to gauge transformations. Thus, in the chiral limit $\hat m=0$, the whole Lagrangian density (\ref{lag}) is gauge symmetric. As a direct consequence of the gauge invariance of four-quark interactions, the following expressions in (\ref{smat3}) must be invariant too
\begin{eqnarray}
\label{sc}
&&\delta \left[ \bar q(\bar s+i\gamma_5 p) q \right]=0, \\
&&\delta \left[ \bar q\gamma^\mu (v_\mu+\gamma_5 a_\mu )q\right]=0.
\end{eqnarray}
Here and in the following we use the notation $\bar s=s-m$.

Let us consider the consequences of the first equation (\ref{sc}). Representing
\begin{equation}
\bar s+i\gamma_5p = P_L (\bar s-ip) + P_R (\bar s+ip),
\end{equation}
and observing that $P_LP_R=0$, one can conclude that the left and right components should be invariant independently, i.e.,
\begin{equation}
\delta \left[ \bar q_R(\bar s-ip) q_L \right]=0, \quad \delta \left[ \bar q_L(\bar s+ip) q_R \right]=0.
\end{equation}

It follows then that
\begin{eqnarray}
\label{gts}
&&\delta \bar s = i [\theta , \bar s]+\{ \beta , p \} , \\
\label{gtp}
&&\delta p = i [\theta , p]-\{\beta , \bar s \},
\end{eqnarray}
where the new set of local parameters has been introduced
\begin{eqnarray}
&& \theta =\frac{1}{2}\left(\omega + e\alpha T_3 \right), \\
&& \beta =-\frac{1}{2}\left(\omega -e\alpha T_3  \right).
\end{eqnarray}

The transformation laws for the spin-1 fields can be obtained in a similar way yielding
\begin{eqnarray}
\label{v}
&&\delta  v_\mu = i [\theta , v_\mu ]+ i [\beta , a_\mu ], \\
\label{a}
&&\delta a_\mu = i [\theta , a_\mu ]+i [\beta , v_\mu ].
\end{eqnarray}

In these notations the gauge transformations remind us of the usual chiral laws, but with local parameters.

The lesson taught by these calculations is that the gauge covariant structure differs from the chiral one through the replacement of usual derivatives of spin-0 fields by the covariant ones. Because the chiral covariant replacement (\ref{nongcov}) contains the derivatives of the pseudoscalar and scalar fields, one has to learn how to construct covariant derivatives of the spin-0 fields under the action of the $SU(2)_L\times U(1)_R$ gauge group.

\section{Gauge Covariant Derivatives of Spin-0 Fields}
\label{gcd}

Having at hand the meson fields which transform under a gauge $SU(2)_L\times U(1)_R$ group in a definite way, we proceed to the second step. Our task now is to derive the gauge covariant derivatives of pseudoscalar and scalar fields. We can find them by using the known quark content of the variables $p_a$ and $s_a$, introduced above as a result of the bosonization. The details can be found, for instance, in \cite{Osipov17jl,Osipov17aph}.

The covariant derivative of the pseudoscalar field $p_a\propto i\bar q\gamma_5\tau_a q$ can be derived by using our knowledge of the covariant derivative of the quark field. This gives
\begin{eqnarray}
{\cal D}_\mu p_a&=&i {\cal D}_\mu \bar q \gamma_5\tau_a q +i\bar q\gamma_5\tau_a{\cal D}_\mu q  \nonumber \\
    &=&i {\cal D}_\mu \bar q_L\tau_a q_R + i\bar q_L\tau_a {\cal D}_\mu q_R \nonumber \\
    &-&i {\cal D}_\mu \bar q_R\tau_a q_L - i\bar q_R\tau_a {\cal D}_\mu q_L \nonumber \\
    &=& \partial_\mu p_a -g(\bar q_LA_\mu\tau_a q_R + \bar q_R\tau_a A_\mu q_L ) \nonumber \\
    &+& g'B_\mu (\bar q_L\tau_aT_3q_R - \bar q_R T_3\tau_a q_L ) \nonumber \\
    &=& \partial_\mu p_a - \frac{g}{2}\bar q \left(\{A_\mu,\tau_a\}+\gamma_5[A_\mu, \tau_a] \right)q\nonumber \\
    &+&\, \frac{g'}{2}B_\mu\bar q\left(\{T_3, \tau_a\}-\gamma_5[T_3, \tau_a] \right)q   \nonumber \\
    &=&\partial_\mu p_a -\frac{1}{2}\bar q\left( \{gA_\mu -g'B_\mu T_3, \tau_a \} \right. \nonumber \\
    &+&\left. \gamma_5[gA_\mu+g'B_\mu T_3, \tau_a]\right)q.
\end{eqnarray}

This result can be put in the more suggestive form
\begin{equation}
{\cal D}_\mu p = {\cal D}_\mu p_a \tau_a =\partial_\mu p-i [N_\mu, p]-\{K_\mu, \bar s \},
\end{equation}
where we have introduced the auxiliary fields
\begin{eqnarray}
&& N_\mu=\frac{1}{2}\left(gA_\mu+g'B_\mu T_3\right), \\
&&K_\mu =\frac{1}{2}\left(gA_\mu -g'B_\mu T_3\right).
\end{eqnarray}

In terms of physical gauge variables, one finds
\begin{eqnarray}
&& {\cal D}_\mu p=\partial_\mu p - \frac{ig}{2}\left[T_+W_\mu^+ + T_-W_\mu^- +T_3 \frac{\cos 2\theta_W}{\cos\theta_W} Z_\mu , p \right] \nonumber \\
&& -ie {\cal A}_\mu [Q,p]
      -\frac{g}{2}\left\{T_+W_\mu^+ + T_-W_\mu^- +\frac{ T_3Z_\mu}{\cos\theta_W} ,\bar s \right\}.
\end{eqnarray}

The covariant derivative of the scalar field can be obtained in a similar way. The result is
 \begin{equation}
{\cal D}_\mu \bar s =\partial_\mu \bar s   -i [N_\mu, \bar s]+\{K_\mu, p \}.
\end{equation}

Both derivatives are gauge covariant, because it is easy to show that they transform like the ordinary pseudoscalar (\ref{gtp}) and scalar (\ref{gts}) fields under $SU(2)_L\times U(1)_R$
\begin{eqnarray}
&& \delta {\cal D}_\mu p= i [\theta , {\cal D}_\mu p] -\{ \beta, {\cal D}_\mu \bar s\}, \\
&& \delta {\cal D}_\mu \bar s= i [\theta , {\cal D}_\mu \bar s] +\{ \beta, {\cal D}_\mu p\}.
\end{eqnarray}
In order to show this it is useful to consider the transformation laws of $N_\mu$ and $K_\mu$ under the action of the gauge group
\begin{eqnarray}
&& \delta N_\mu =\frac{i}{2} [\omega, N_\mu +K_\mu]+ \partial_\mu\theta, \\
&& \delta K_\mu =\frac{i}{2} [\omega, N_\mu +K_\mu]- \partial_\mu\beta.
\end{eqnarray}

In summary: gauge covariant derivatives have the structure of chiral-covariant derivatives. Thus, if in the chiral-invariant expression we replace the usual derivatives by the gauge covariant ones, the result will be gauge and chiral invariant.

\section{Gauge Covariant Diagonalization}
\label{cd}

Let us recall now, that the chiral-covariant version of the replacement (\ref{noncd}) has been worked out in \cite{Osipov00}. It has been shown there that the covariant redefinition of the axial-vector field cannot be done without a corresponding change in its chiral partner, i.e., the vector field. This is a direct consequence of the linear realization of chiral symmetry. As a result, it has been found that two bilinear combinations of scalar and pseudoscalar fields transform like axial-vector and vector fields and are chiral partners. We will use this result, but to extend it to be covariant under transformations of the gauge group, we replace the usual derivatives of spin-0 fields by the gauge covariant ones. Thus, to avoid the $p-a_\mu$ mixing,  instead of (\ref{nongcov}) we will use the following replacement
\begin{eqnarray}
\label{da}
&& a_\mu\to a_\mu +\frac{\kappa}{2}\,Y_\mu , \\
\label{dv}
&& v_\mu\to v_\mu +\frac{\kappa}{2}\,X_\mu,
\end{eqnarray}
where
\begin{eqnarray}
\label{xmu}
&& X_\mu = i\left(   [p,{\cal D}_\mu p]+[\bar s, {\cal D}_\mu\bar s] \right), \\
\label{ymu}
&& Y_\mu = \{p, {\cal D}_\mu\bar s \}-\{\bar s, {\cal D}_\mu p \}.
\end{eqnarray}
One can check that $X_\mu^\dagger =X_\mu$ and $Y_\mu^\dagger =Y_\mu$.

The coupling of $X_\mu$ and $Y_\mu$ with quarks has a gauge and a chiral invariant form. Indeed, let us show the gauge covariant property of these chiral partners. For that we consider first how $X_\mu$ transforms under the action of the gauge $SU(2)_L\times U(1)_R$ group:
\begin{eqnarray}
\delta X_\mu = &i& [i[\theta , p] -\{\beta , \bar s\}, {\cal D}_\mu p]  \nonumber \\
+&i&[p, i[\theta, {\cal D}_\mu p]-\{\beta , {\cal D}_\mu \bar s\} ] \nonumber\\
 +&i&[i[\theta,\bar s]+\{\beta, p\}, {\cal D}_\mu\bar s] \nonumber \\
 +&i&[\bar s, i[\theta, {\cal D}_\mu \bar s]+\{\beta, {\cal D}_\mu p\}].
\end{eqnarray}
Now, we use the various Jacobi identities
\begin{eqnarray}
&& [[\theta ,p],{\cal D}_\mu p ] + [p, [\theta , {\cal D}_\mu p ]]=[\theta , [p, {\cal D}_\mu p]], \\
&& [[\theta ,\bar s],{\cal D}_\mu \bar s ] + [\bar s, [\theta , {\cal D}_\mu \bar s ]]=[\theta , [\bar s, {\cal D}_\mu \bar s]], \\
&& [ \{\beta, \bar s\} ,{\cal D}_\mu p ] - [\bar s, \{\beta, {\cal D}_\mu p  \}]=[\beta, \{\bar s, {\cal D}_\mu p \}], \\
&& [p, \{\beta , {\cal D}_\mu \bar s  \}]-[\{\beta , p\}, {\cal D}_\mu \bar s ]=-[\beta, \{p, {\cal D}_\mu \bar s\}].
\end{eqnarray}
This gives
\begin{equation}
\delta X_\mu =i [\theta, X_\mu ]+i[\beta, Y_\mu].
\end{equation}
That agrees well with (\ref{v}).

Let us consider now the gauge transformation of $Y_\mu$
\begin{eqnarray}
\delta Y_\mu =&-&\{i[\theta, \bar s]+\{\beta, p\}, {\cal D}_\mu p\}  \nonumber \\
                       &-&\{\bar s, i[\theta, {\cal D}_\mu p] -\{\beta, {\cal D}_\mu \bar s\}\} \nonumber \\
                       &+&\{i[\theta, p] -\{\beta, \bar s \}, {\cal D}_\mu \bar s\}   \nonumber \\
                       &+&\{p, i[\theta, {\cal D}_\mu \bar s]+\{\beta, {\cal D}_\mu p\}\}.
\end{eqnarray}
Also, applying the Jacobi identities
\begin{eqnarray}
&&\{[\theta, \bar s], {\cal D}_\mu p \}+\{\bar s, [\theta, {\cal D}_\mu p] \}=[\theta, \{\bar s,{\cal D}_\mu p\}],      \\
&&\{[\theta, p], {\cal D}_\mu \bar s \}+\{p, [\theta, {\cal D}_\mu \bar s] \}=[\theta, \{p,{\cal D}_\mu \bar s\}],      \\
&& \{\{\beta, p\}, {\cal D}_\mu p \} -\{p, \{\beta, {\cal D}_\mu p\}\}=[\beta, [p, {\cal D}_\mu p]], \\
&& \{\{\beta, \bar s\}, {\cal D}_\mu \bar s \} -\{\bar s, \{\beta, {\cal D}_\mu \bar s\}\}=[\beta, [\bar s, {\cal D}_\mu \bar s]],
\end{eqnarray}
we find
\begin{equation}
\delta Y_\mu = i [\theta, Y_\mu ]+i[\beta, X_\mu].
\end{equation}
This is precisely the transformation law satisfied by the axial-vector field (\ref{a}).

The diagonalization (\ref{da}) and (\ref{dv}) has the following  consequences. First, it changes the form of the Dirac operator by introducing a set of linear and nonlinear interactions of mesons and gauge bosons with quarks
\begin{eqnarray}
D_m\to D_m&=&i\gamma^\mu{\cal D}_\mu +\bar s+i\gamma_5p +\gamma^\mu (v_\mu +\gamma_5 a_\mu) \nonumber \\
&+&\frac{\kappa}{2}\gamma^\mu (X_\mu + \gamma_5 Y_\mu).
\end{eqnarray}

The latter can be rewritten in a more convenient way as follows
\begin{equation}
D_m=i\gamma^\mu\partial_\mu +\bar s +i\gamma_5 p +\gamma^\mu\Gamma_\mu,
\end{equation}
where
\begin{equation}
\Gamma_\mu =\Gamma_\mu^{(V)}+\gamma_5\Gamma_\mu^{(A)}
\end{equation}
and
\begin{eqnarray}
\Gamma_\mu^{(V)}&=&v_\mu+eQ{\cal A}_\mu +\frac{gZ_\mu}{2\cos\theta_W}\left(T_3-Q\sin^2\theta_W\right) \nonumber \\
&+& \frac{g}{2}\left(T_+W_\mu^++T_-W_\mu^-\right) +\frac{\kappa}{2}\,X_\mu ,\\
\Gamma_\mu^{(A)}&\!\!\!=\!\!\!&a_\mu-\frac{gT_3Z_\mu}{2\cos\theta_W} -\frac{g}{2}\left(T_+W_\mu^++T_-W_\mu^-\right) \nonumber \\
&+& \frac{\kappa}{2}\,Y_\mu .
\end{eqnarray}

Second, after replacements (\ref{da}) and (\ref{dv}), the Lagrangian density (\ref{Mpart1}) gets some vertices with interactions. We will consider this point later on in detail.

\section{Vector and Axial-Vector Meson Dominance}
\label{vavd}
The vector and axial-vector meson dominance is a natural consequence of the extended NJL model, provided that the $p-a_\mu$ diagonalization is made in accord with Eq.(\ref{noncd}). The covariant diagonalization adds to this picture a set of new interactions while keeping the old results unchanged. To show this let us eliminate the Yukawa interactions of the photon, $Z$ and $W^\pm$ bosons with quarks.

In the case of the vector fields, we have to avoid only the linear interactions of ${\cal A}_\mu, Z_\mu$ and $W^\pm_\mu$ induced by the covariant derivative of quarks ${\cal D}_\mu q$. The vector $X_\mu$ is quadratic [at least] in boson fields and, therefore, does not contain such terms. We see that the replacement
\begin{eqnarray}
\label{vdr}
v_\mu&\to& v_\mu -eQ{\cal A}_\mu -\frac{gZ_\mu}{2\cos\theta_W}\left(T_3-Q\sin^2\theta_W\right) \nonumber\\
&-&\frac{g}{2} \left(T_+W_\mu^++T_-W_\mu^-\right)
\end{eqnarray}
leads to the desired result. This replacement has the following isospin content
\begin{eqnarray}
&&v_\mu^0\to v_\mu^0 -\frac{e}{6} {\cal A}_\mu +\frac{gZ_\mu \sin^2\theta_W}{12\cos\theta_W}, \nonumber \\
&& v_\mu^3\to v_\mu^3-\frac{e}{2}{\cal A}_\mu -\frac{g}{4}\cos\theta_WZ_\mu, \nonumber \\
&& v_\mu^\pm\to v_\mu^\pm - \frac{g}{4}W_\mu^\pm .
\end{eqnarray}

In the axial-vector case, the combination $Y_\mu$ contains a linear term with $Z_\mu$ and $W^\pm_\mu$. This is the second term in the following expression
\begin{eqnarray}
Y_\mu &=& 2m\partial_\mu p +2gm^2\left(\frac{T_3Z_\mu}{\cos\theta_W}+T_+W_\mu^++T_-W_\mu^-\right) \nonumber \\
&+& {\cal O}( \mbox{field}^2 ).
\end{eqnarray}
It should be subtracted by the appropriate redefinition of the axial-vector field together with the corresponding terms from ${\cal D}_\mu q$. The replacement
\begin{equation}
\label{avd}
a_\mu\to a_\mu +g_A\, \frac{g}{2}\left(\frac{T_3Z_\mu}{\cos\theta_W} +T_+W_\mu^++T_-W_\mu^-\right),
\end{equation}
where $g_A=1-2\kappa m^2$, yields  the desired result. In components, it has the following content
\begin{eqnarray}
&& a_\mu^0\to a_\mu^0, \nonumber \\
&& a_\mu^3\to a_\mu^3+ \frac{gg_AZ_\mu}{4\cos\theta_W}, \nonumber \\
&& a_\mu^\pm \to a_\mu^\pm +g_A\,\frac{g}{4} W_\mu^\pm.
\end{eqnarray}

These replacements change $\Gamma_\mu $ accordingly
\begin{eqnarray}
\label{gammav}
\Gamma_\mu^{(V)}&\to&\Gamma_\mu^{(V)}=v_\mu +\frac{\kappa}{2}\,X_\mu ,\\
\label{gammaa}
\Gamma_\mu^{(A)}&\to&\Gamma_\mu^{(A)}=a_\mu +\frac{\kappa}{2}\,Y_\mu \nonumber \\
     &-& \kappa gm^2\left(\frac{T_3Z_\mu}{\cos\theta_W}+T_+W_\mu^++T_-W_\mu^-\right).
\end{eqnarray}
One can see that $\Gamma_\mu$ still contains the gauge fields inside the covariant derivatives in $X_\mu$ and $Y_\mu$. This is a signal that the theory may lead to some deviations from the vector and axial-vector dominance, because $\Gamma_\mu$ is directly coupled to quarks. 

We obtain that similar changes occur in the original mass term of spin-1 fields (\ref{Mpart1}) where the above redefinitions induce the replacements
\begin{eqnarray}
 v_\mu &\to& \tilde v_\mu = v_\mu +\frac{\kappa}{2} X_\mu -eQ{\cal A}_\mu \nonumber\\
            &-& \frac{gZ_\mu}{2\cos\theta_W}\left(T_3-Q\sin^2\theta_W\right) \nonumber \\
            &-&\frac{g}{2}\left(T_+W_\mu^++T_-W_\mu^-\right) , \nonumber \\
a_\mu &\to& \tilde a_\mu = a_\mu +\frac{\kappa}{2}Y_\mu \nonumber\\
&+& g_A \frac{g}{2}\left(\frac{T_3Z_\mu}{\cos\theta_W} +T_+W_\mu^++T_-W_\mu^-\right).
\end{eqnarray}
From that we have
\begin{eqnarray}
\label{vt2}
\mbox{tr}&&\left(\tilde v_\mu^2\right) = \mbox{tr}\left[\left(v_\mu + \frac{\kappa}{2}\, \xi_\mu \right)  \left(v_\mu + \frac{\kappa}{2} \,\xi_\mu +\kappa\,\Xi_\mu \right) \right] \nonumber \\
&& -2e{\cal A}_\mu \left(\frac{v_{\mu 0}}{3}+v_{\mu 3} +  \frac{\kappa}{2} \,\xi_{\mu 3} \right)\nonumber\\
&&+\frac{gZ_\mu}{\cos\theta_W}\left[ \frac{v_{\mu 0}}{3}\sin^2\theta_W - \left( v_{\mu 3} + \frac{\kappa}{2}\,\xi_{\mu 3}\right)\cos^2\theta_W  \right] \nonumber \\
&& -g\left[ W_\mu^+  \left( v_{\mu}^- + \frac{\kappa}{2}\,\xi_{\mu -}\right) +
                 W_\mu^-  \left( v_{\mu}^+ + \frac{\kappa}{2}\,\xi_{\mu +}\right) \right] \nonumber \\
&& +\left( \kappa\, \Xi_{\mu +} - \frac{g}{2}\,W_\mu^+ \right)\left( \kappa\, \Xi_{\mu -} - \frac{g}{2}\,W_\mu^- \right) \nonumber\\
&&+\frac{1}{18}\left( \frac{gZ_\mu}{2\cos\theta_W}\sin^2\theta_W - e {\cal A}_\mu  \right)^2 \nonumber \\
&&+\frac{1}{2} \left( \kappa\,\Xi_3 - e {\cal A}_\mu -\frac{g}{2} \, Z_\mu \cos\theta_W  \right)^2 ,
\end{eqnarray}
and
\begin{eqnarray}
\label{at2}
\mbox{tr}&&\left(\tilde a_\mu^2\right) = \mbox{tr}\left(a_\mu +\frac{\kappa}{2}\, Y_\mu\right)^2 +
\frac{gg_A Z_\mu}{\cos\theta_W} \left( a_{\mu  3} +\frac{\kappa}{2}\, Y_{\mu 3}  \right)    \nonumber\\
&&+gg_A \left[ W_\mu^+  \left( a_{\mu}^- + \frac{\kappa}{2}\, Y_{\mu -}\right) +
                 W_\mu^-  \left( a_{\mu}^+ + \frac{\kappa}{2}\, Y_{\mu +}\right) \right] \nonumber \\
&&+\frac{1}{4}\,g^2g_A^2 \left(W_\mu^+ W_\mu^- +\frac{Z_\mu^2}{2\cos^2\theta_W}   \right).
\end{eqnarray}
Here we use the following conventions [additionally to Eqs. (\ref{xmu}) and (\ref{ymu})]:
\begin{eqnarray}
X_\mu &=& \xi_\mu + \Xi_\mu , \quad \xi_\mu =i\left([p,\partial_\mu p]+[s, \partial_\mu s]\right), \nonumber\\
Y_\mu &=&\zeta_\mu + \Psi_\mu, \quad \zeta_\mu= \{p, \partial_\mu \bar s \}-\{\bar s, \partial_\mu  p \}.
\end{eqnarray}
It is assumed that these four-vectors are imbedded into the Lie algebra of the $U(2)$ group as follows $\zeta_\mu=\zeta_{\mu a}\tau_a$. We emphasize that the second order electroweak contributions $\sim e^2$, $eg$ or $g^2$ play an important role to protect the gauge symmetry of the vector dominance approach. A simple example showing the details of such mechanism has been considered by Sakurai in \cite{Sakurai69}.

Let us extract from (\ref{vt2})  and (\ref{at2}) only the terms of the second order in powers of fields. For that, from $\zeta_\mu$'s and $\Psi_\mu$'s one should keep the following terms
\begin{eqnarray}
\label{expan}
&&\zeta_{\mu 0}=2m\partial_\mu p_0 + {\cal O}(\mbox{field}^2), \nonumber \\
&&\zeta_{\mu 3}=2m\partial_\mu p_3 + {\cal O}( \mbox{field}^2), \nonumber \\
&&\zeta_{\mu \pm}=2m\partial_\mu p^\pm + {\cal O}( \mbox{field}^2), \nonumber \\
&&\Psi_{\mu 3}=gm^2 \frac{Z_\mu}{\cos\theta_W} + {\cal O}(\mbox{field}^2), \nonumber \\
&&\Psi_{\mu \pm} =gm^2 W_\mu^\pm +{\cal O} (\mbox{field}^2) .
\end{eqnarray}
Combining  (\ref{vt2}), (\ref{at2}) with (\ref{expan}) we find that
\begin{eqnarray}
{\cal L}_{M} &\to& \frac{1}{4G_V}\left\{
\mbox{tr}[v_\mu^2+(a_\mu + \kappa m \partial_\mu p)^2] \right. \nonumber\\
&-&2e{\cal A}_\mu \left(\frac{v_{\mu}^0}{3}+v_\mu^3 \right) +\frac{gZ_\mu}{\cos\theta_W} \nonumber \\
&\times&\left(\frac{v_\mu^0}{3} \sin^2\theta_W - v_\mu^3 \cos^2\theta_W + a_\mu^3 +\kappa m\partial_\mu p_3 \right)  \nonumber\\
&-& g\left[W_\mu^+\left(v_\mu^- -a_\mu^--\kappa m \partial_\mu p^-\right)+\mbox{h.c.}\right]  \nonumber \\
&+&\frac{5}{9}e^2{\cal A}_\mu^2  + \frac{g^2}{4}W_\mu^+W_\mu^- \nonumber \\
&+& \frac{g^2Z_\mu^2}{8\cos^2\theta_W}\left(1+\cos^4\theta_W +\frac{1}{9}\sin^4\theta_W  \right) \nonumber \\
&+&\left. \frac{e}{2}g Z_\mu {\cal A}_\mu \left(\cos\theta_W-\frac{\sin^2\theta_W}{9 \cos\theta_W} \right) \right\}.
\end{eqnarray}
Or, after the standard redefinition of the meson fields [see Eqs. (\ref{Rsp}) - (\ref{Rvach})], we arrive at the conventional vector and axial-vector meson dominance results
\begin{equation}
\label{vmd}
{\cal L}_\gamma =-e\frac{m_\rho^2}{g_\rho}{\cal A}_\mu\left(\frac{\omega_\mu}{3}+\rho^0_\mu \right),
\end{equation}
\begin{equation}
\label{wd}
{\cal L}_{W^\pm} =\frac{g}{2}W_\mu^\pm\left[\frac{m_\rho^2}{g_\rho} \left(a_{1\mu}^\mp -\rho_\mu^\mp \right)+f_\pi\partial_\mu \pi^\mp \right],
\end{equation}
including mixing of the $Z$ boson with the neutral vector, axial-vector mesons and the pion
\begin{eqnarray}
\label{zd}
{\cal L}_Z&=&\frac{gZ_\mu}{2\cos\theta_W}\left[\frac{m_\rho^2}{g_\rho} \left(\frac{\omega_\mu}{3}\sin^2\theta_W-\rho^0_\mu \cos^2\theta_W +a_{1\mu}^0 \right) \right. \nonumber\\
&+&\left. f_\pi \partial_\mu\pi^0 \right].
\end{eqnarray}
Eq. (\ref{vmd}) summarizes the $U(2)$ field-current identities, describing electromagnetic VMD phenomena. Eq. (\ref{wd}) represents the corresponding result for the charged hadronic weak current. Eq. (\ref{zd}) shows weak mixing of neutral spin-1 states with the $Z$ boson. 

Thus, in the covariant approach, the theory contains both the vertices of the standard vector and axial-vector dominance picture (\ref{vmd})-(\ref{zd}), and the vertices of the non VMD type, originated through the $\Gamma_\mu$ term (\ref{gammav})-(\ref{gammaa}) in the effective quark-meson Lagrangian.

\section{Schwinger-DeWitt Expansion}
\label{sdwe}

Our purpose now is to obtain the effective meson theory defined by the vacuum-to-vacuum transition amplitude (\ref{smat3}), where the operator $D_m$, after all replacements, is given by
\begin{equation}
D_m=i\gamma^\mu \partial_\mu +\bar s +i\gamma_5 p + \gamma^\mu \left( \Gamma_\mu^{(V)}+\gamma_5\Gamma_\mu^{(A)}\right).
\end{equation}
Here, $\Gamma_\mu^{(V)}$ and $\Gamma_\mu^{(A)}$ are given by Eqs. (\ref{gammav}) and (\ref{gammaa}).

Let us integrate in (\ref{smat3}) over the quark fields
\begin{eqnarray}
\label{qint}
\int && [dq] [d\bar q] \exp\left( {i\!\!\int\!\! d^4x\bar qD_mq}\right)=\det D_m \nonumber \\
&&=\mbox{exp }\left( \mbox{Tr} \ln D_m \right).
\end{eqnarray}
The Gaussian path integral accounts for the one-quark-loop contribution to the effective action. The result is given by the non-local functional determinant (up to an overall constant). The trace "Tr" should be calculated over color, Dirac, flavor indices and it also includes the integration over coordinates of the Minkowski space-time.

In particular, the contribution of the chiral determinant to the real part of effective action is
\begin{equation}
\label{seff}
S_{SD}= -\frac{i}{2}\,\mbox{Tr}\ln D_m^\dagger D_m =i\mathcal L_{SD}.
\end{equation}
The consistent approximation scheme to obtain from the non-local chiral determinant (\ref{qint}) the local long wavelength (low-energy) expansion for the effective action $S_{SD}$ is the Schwinger-DeWitt technique \cite{Schwinger54,DeWitt65,Ball89} [see details, for instance, in \cite{Osipov17aph}]. We will restrict ourselves to the first and second-order Seeley-DeWitt coefficients. These coefficients accumulate the divergent part of the effective action, which is regularized here by a covariant ultraviolet cutoff $\Lambda$. Let us recall that the result of such calculations is well known [in the sense that the only difference between the expression for $D_m$ obtained in \cite{Osipov17aph} and $D_m$ here is the replacement of the vector $v_\mu$ and axial-vector field $a_\mu$ by $\Gamma_\mu^{(V)}$ and $\Gamma_\mu^{(A)}$ correspondingly]. Thus, we can use that result by writing
\begin{eqnarray}
\label{LSD1}
{\cal L}_{SD}&=&I_2\,\mbox{tr}\left\{ (\bigtriangledown_\mu s)^2 +  (\bigtriangledown_\mu p )^2  - (s^2-2m s+p^2)^2 \right. \nonumber \\
&-&\left. \frac{1}{3}(v_{\mu\nu} ^2 +a_{\mu\nu}^2)\right\},
\end{eqnarray}
where the factor $I_2$ is given by
\begin{equation}
\label{j01}
I_2 =\frac{N_c}{(4\pi )^2}\left[\ln\left(1+\frac{\Lambda^2}{m^2}\right)-\frac{\Lambda^2}{\Lambda^2+m^2}\right],
\end{equation}
with $N_c$ being the number of color degrees of freedom, and
\begin{eqnarray}
v_{\mu\nu}&=& \partial_\mu \Gamma_\nu^{(V)}-\partial_\nu \Gamma_\mu^{(V)}-i[\Gamma_\mu^{(V)},\Gamma_\nu^{(V)}]-i[\Gamma_\mu^{(A)},\Gamma_\nu^{(A)}], \nonumber\\
a_{\mu\nu}&=& \partial_\mu \Gamma_\nu^{(A)}-\partial_\nu \Gamma_\mu^{(A)}-i[\Gamma_\mu^{(V)},\Gamma_\nu^{(A)}]+i[\Gamma_\nu^{(V)},\Gamma_\mu^{(A)}], \nonumber \\
\bigtriangledown_\mu s&=&\partial_\mu s -i[\Gamma_\mu^{(V)}, s]-\{ \Gamma_\mu^{(A)} ,p\}, \nonumber \\
\bigtriangledown_\mu p&=& \partial_\mu p -i [\Gamma_\mu^{(V)} ,p]+\{\Gamma_\mu^{(A)},\bar s \}.
\end{eqnarray}

The total effective Lagrangian density is the following sum
\begin{equation}
\label{Ltot}
{\cal L}_{tot}={\cal L}_1+ {\cal L}_2 +{\cal L}_{SD}+{\cal L}_{EW},
\end{equation}
where ${\cal L}_{SD}$ and ${\cal L}_{EW}$ are given by Eqs. (\ref{LSD1}) and (\ref{lagem}) correspondingly, and
\begin{eqnarray}
&& {\cal L}_1 = - \frac{\hat m }{4mG_S}   \mbox{tr}   \left(s^2+p^2\right), \nonumber \\
&& {\cal L}_2 = \frac{1}{4G_V}\mbox{tr}\left(\tilde v_\mu^2+\tilde a_\mu^2\right).
\end{eqnarray}

Some comments about formula (\ref{Ltot}) are still in order. To get this Lagrangian density we have used the gap equation
\begin{equation}
\label{gap}
m-\hat m= mG_S I_1,
\end{equation}
where
\begin{equation}
I_1=\frac{N_c}{2\pi^2}\left[\Lambda^2 -m^2\ln \left(1+\frac{\Lambda^2}{m^2}\right)\right].
\end{equation}
It is assumed that the strength of the quark interactions is large enough, $G_S > (2\pi )^2/(N_c\Lambda^2)$, to generate a non-trivial, $m \neq 0$, solution of Eq. (\ref{gap}) [even if the current quarks would be massless].

The Lagrangian density ${\cal L}_{tot}$ does not contain $p-a_\mu$ mixing. This is because of the cancellation which occurs between the following two different contributions to the non diagonal $p-a_\mu$ mixing term in ${\cal L}_{tot}$. That restrict the numerical value of the parameter $\kappa$ to 
\begin{equation}
\label{pac}
\frac{1}{2\kappa} = m^2 +\frac{1}{16G_V I_2}.
\end{equation}

Indeed, from ${\cal L}_2$ one finds
\begin{equation}
{\cal L}_2= \frac{2\kappa m}{4G_V}\,\mbox{tr}\left(a_\mu\partial_\mu p\right)+ \ldots
\end{equation}
The other contribution is induced by ${\cal L}_{SD}$. The term with $(\bigtriangledown_\mu p )^2$ is responsible for it
\begin{eqnarray}
{\cal L}_{SD}&=& I_2 \mbox{tr}(\bigtriangledown_\mu p )^2+\ldots = I_2 \mbox{tr} \left( g_A\partial_\mu p -2m a_\mu \right)^2 +\ldots  \nonumber \\
&=&-4mg_A I_2 \mbox{tr}\left(a_\mu\partial_\mu p\right)+ \ldots
\end{eqnarray}
As a result one obtains
\begin{equation}
{\cal L}_1+{\cal L}_{SD} \to 2m\left(\frac{\kappa}{4G_V}-2 g_A I_2 \right)\mbox{tr}
\left(a_\mu\partial_\mu p\right) =0
\end{equation}
This sum can be zero only if the expression in parenthesis is zero. This yields (\ref{pac}).

The free part of the Lagrangian density ${\mathcal L}_{tot}$ must display a canonical form. This can be done by the redefinition of the fields
\begin{eqnarray}
\label{Rsp}
&& s =g_\sigma \sigma, \qquad  \ \   p=g_\pi \pi, \\
\label{Rvan}
&& v^0_\mu =\frac{g_\rho}{2}\omega_\mu, \quad  a^0_\mu =\frac{g_\rho}{2} f_{1\mu}, \\
\label{Rvach}
&&\vec v_\mu=\frac{g_\rho}{2}\vec\rho_\mu, \quad \vec a_\mu =\frac{g_\rho}{2}\vec a_{1\mu}.
\end{eqnarray}
The effective constants $g_\sigma , g_\pi , g_\rho$ and masses of meson states are functions of the $I_2$ and the constant $Z=g_A^{-1}$
\begin{eqnarray}
\label{couplings}
&&g_\sigma^2=\frac{1}{4I_2},\quad  g_\pi^2=Zg_\sigma^2,\quad  g_\rho^2=6g_\sigma^2, \\
&&m_\pi^2=\frac{\hat mg_\pi^2}{mG_S},\quad m_\sigma^2=4m^2+g_A m_\pi^2, \\
\label{mrho}
&&m_\rho^2=m_\omega^2=\frac{3}{8G_VI_2}, \\
\label{mrel}
&&m_{a_1}^2=m_{f_1}^2=m_\rho^2+6m^2.
\end{eqnarray}

To complete this section let us present ${\cal L}_{SD}$ in a form where the electroweak part of the Schwinger-DeWitt contribution is explicitly separated from the pure strong interactions of mesons. This gives us a clear picture of the contributions to the weak hadronic current beyond the vector and axial-vector meson dominance. We will extract only the leading order terms in powers of the electroweak couplings. After some algebraic manipulations we obtain
 \begin{eqnarray}
 \label{LSD}
{\cal L}_{SD}&=&I_2\, \mbox{tr}\left\{ (\overline{\bigtriangledown_\mu s})^2
+  (\overline{\bigtriangledown_\mu p} )^2  - (s^2-2m s+p^2)^2 \right. \nonumber \\
&-& \frac{1}{3}(\overline{v}_{\mu\nu} ^2 +\overline{a}_{\mu\nu}^2)
-\kappa \overline{\bigtriangledown_\mu s} \left( \{\overline{\Psi}_\mu, p\}+i [\Xi_\mu, s] \right) \nonumber\\
&+& \kappa \overline{\bigtriangledown_\mu p} \left( \{\overline{\Psi}_\mu, \bar s\}-i [\Xi_\mu, p] \right) \nonumber\\
&+&\frac{2\kappa}{3} \, \overline{v}_{\mu\nu} \left(\partial_\nu \Xi_\mu +i [\Xi_\mu,  v_\nu+\frac{\kappa}{2}\xi_\nu] \right.  \nonumber\\
&+&\left.  i [\overline{\Psi}_\mu,  a_\nu
      +\frac{\kappa}{2}\zeta_\nu ] \right) + \frac{2\kappa}{3} \overline{a}_{\mu\nu} \left(\partial_\nu \overline{\Psi}_\mu \right. \nonumber\\
&+&\left.\left.i [\Xi_\mu,  a_\nu+\frac{\kappa}{2}\zeta_\nu] + i [\overline{\Psi}_\mu, v_\nu+\frac{\kappa}{2}\xi_\nu ]\right)\right\} \nonumber \\
&+& {\cal O}(g^2).
\end{eqnarray}
Here, only the first four terms represent the pure strong interactions of mesons. The terms proportional to $\kappa $ describe the interactions of gauge bosons with clusters of mesons. So, the corresponding hadronic electroweak current is a non linear function in meson fields. In (\ref{LSD}) we show only the terms which are of order $g$ in electroweak interactions. The higher order terms are not the goal of the present work, but they should be considered, for instance, if one would study two photon interactions, or $\tau$-decays into mesons with one or more photons in the final state. We use the following notations for the pure hadron contributions
\begin{eqnarray}
\overline{\bigtriangledown_\mu s}&=& \partial_\mu s - i[v_\mu+\frac{\kappa}{2}\xi_\mu , s]- \{ a_\mu+ \frac{\kappa}{2}\zeta_\mu ,p\}, \\
\overline{\bigtriangledown_\mu p}&=& \partial_\mu p -i[v_\mu+\frac{\kappa}{2}\xi_\mu , p] + \{ a_\mu+ \frac{\kappa}{2}\zeta_\mu ,\bar s \}, \\
\overline{v}_{\mu\nu}&=& \partial_\mu \left(v_\nu+\frac{\kappa}{2}\xi_\nu\right)
                                   -\partial_\nu \left(v_\mu+\frac{\kappa}{2}\xi_\mu\right)    \nonumber \\
&-&i[v_\mu+\frac{\kappa}{2}\xi_\mu,  v_\nu+\frac{\kappa}{2}\xi_\nu ] \nonumber \\
&-&i [a_\mu+\frac{\kappa}{2}\zeta_\mu,  a_\nu+\frac{\kappa}{2}\zeta_\nu ], \\
\overline{a}_{\mu\nu} &=& \partial_\mu \left(a_\nu +\frac{\kappa}{2}\zeta_\nu\right)
       -\partial_\nu \left(a_\mu +\frac{\kappa}{2}\zeta_\mu\right) \nonumber \\
&-& i [v_\mu + \frac{\kappa}{2}\xi_\mu,  a_\nu +\frac{\kappa}{2}\zeta_\nu]    \nonumber \\
&+& i[v_\nu + \frac{\kappa}{2}\xi_\nu,  a_\mu +\frac{\kappa}{2}\zeta_\mu].
\end{eqnarray}

The $\Xi_\mu$, and $\overline{\Psi}_\mu$ are linear functions in electroweak fields
\begin{equation}
\label{psibar}
\overline{\Psi}_\mu =\Psi_\mu -2 gm^2\left(\frac{T_3Z_\mu}{\cos\theta_W}+T_+W_\mu^++T_-W_\mu^-\right).
\end{equation}

\section{One Photon Electromagnetic Interactions}
\label{ei}

To make our result more transparent, we will present here the electromagnetic part of the entire effective meson Lagrangian density at leading order in $e$, which we put in the standard form
\begin{equation}
\label{Lem}
{\cal L}_{em}=e{\cal A}_\mu j_\mu^{em}.
\end{equation}
The hadronic electromagnetic current, $j_\mu^{em}$, depends on meson fields. For convenience, the current $j_\mu^{em}$ will be presented in the form where we introduce two useful matrices. The first one collects a set of meson fields which interact with the electromagnetic field as it follows from the matrix $\Xi_\mu$, i.e., $\Xi_\mu\to 4e{\cal A}_\mu \Xi^{em}$, where
\begin{equation}
\Xi^{em}=\left( \begin{array}{cc} -(p^+p^- +s^+s^-) & \frac{1}{\sqrt{2}}(p^+p_3+s^+s_3) \\
\frac{1}{\sqrt{2}}(p^-p_3+s^-s_3)  &   (p^+p^- +s^+s^-)   \end{array}\right) .
\end{equation}

The second matrix contains the meson fields which come together with the electromagnetic field from $\Psi_\mu$,i.e., $\Psi_\mu\to 4ie{\cal A}_\mu \Psi^{em}$, where
\begin{equation}
\Psi^{em}=\left( \begin{array}{cc} p^+s^- -p^-s^+ & \frac{1}{\sqrt{2}}(p^+\bar s_0-p_0s^+) \\
\frac{1}{\sqrt{2}}(p_0s^- -p^-\bar s_0)  &   p^+s^- -p^-s^+   \end{array}\right) .
\end{equation}

Using these notations, we can write $j_\mu^{em}$ as follows
\begin{eqnarray}
\label{jem}
j_\mu^{em}&=& \frac{1}{2G_V}\mbox{tr} \left[ 2i\kappa \Psi^{em} \left(a_\mu+\frac{\kappa}{2}\zeta_\mu \right)\right. \nonumber \\
&-&\left. \left(v_\mu +\frac{\kappa}{2} \xi_\mu \right)\left(Q-2\kappa \Xi^{em}\right)\right] \nonumber \\
&+&4I_2\mbox{tr}\left\{  \kappa \overline{\bigtriangledown_\mu p} \left( \{i\Psi^{em}, \bar s\}-i[\Xi^{em},p]  \right)\right. \nonumber \\
&-& \kappa \overline{\bigtriangledown_\mu s} \left( \{i\Psi^{em}, p\}+i[\Xi^{em},s]  \right) \nonumber \\
&+&2i\frac{\kappa}{3}\overline{v}_{\mu\nu}\left( [\Xi^{em},v_\nu + \frac{\kappa}{2}\xi_\nu ]\right. \nonumber\\
&+&\left. [i\Psi^{em}, a_\nu +\frac{\kappa}{2}\zeta_\nu] \right) + 2i\frac{\kappa}{3}\overline{a}_{\mu\nu} \nonumber\\
&\times& \left( [\Xi^{em}, a_\nu +\frac{\kappa}{2}\zeta_\nu ]  + [i\Psi^{em}, v_\nu + \frac{\kappa}{2}\xi_\nu ] \right)\nonumber\\
&-&2\left. \frac{\kappa}{3}\left( \partial_\nu \overline{v}_{\mu\nu} \Xi^{em}+ \partial_\nu \overline{a}_{\mu\nu} i\Psi^{em}  \right)\right\}.
\end{eqnarray}
Although we have established Eq. (\ref{Lem}) only to the first order in $e$, the inclusion of all orders in $e$ is possible, but a full discussion will not be given in this paper.

\subsection{$\gamma\pi\pi$ vertex}

This vertex has been considered in \cite{Osipov18vd} in a slightly different approach. It has been shown that the VMD result remains unchanged when one uses the covariant replacement (\ref{covd}) to describe the electromagnetic form factor of the pion. Our goal here is to show that the current $j_\mu^{em}$ leads to the same conclusion, as it should be.

Indeed, extracting from $\overline{\bigtriangledown_\mu p} $ only the terms relevant for the $\gamma\pi\pi$ amplitude 
\begin{equation}
\overline{\bigtriangledown_\mu p} \to \partial_\mu p -\kappa m \zeta_\mu \to g_A \partial_\mu p,
\end{equation}
we find
\begin{eqnarray}
j_\mu^{em} &\to& \frac{i\kappa^2}{2G_V} \mbox{tr}\left(\Psi^{em}\zeta_\mu\right) - 8i\kappa m g_A I_2 \mbox{tr} \left(\Psi^{em} \partial_\mu p  \right) \nonumber\\
&\to& i\kappa m \left(\frac{\kappa}{G_V}-8g_A I_2 \right) \mbox{tr} \left(\Psi^{em} \partial_\mu p\right)=0.
\end{eqnarray}
The second step is a consequence of Eq. (\ref{pac}). 

\begin{figure}
\resizebox{0.40\textwidth}{!}{%
 \includegraphics{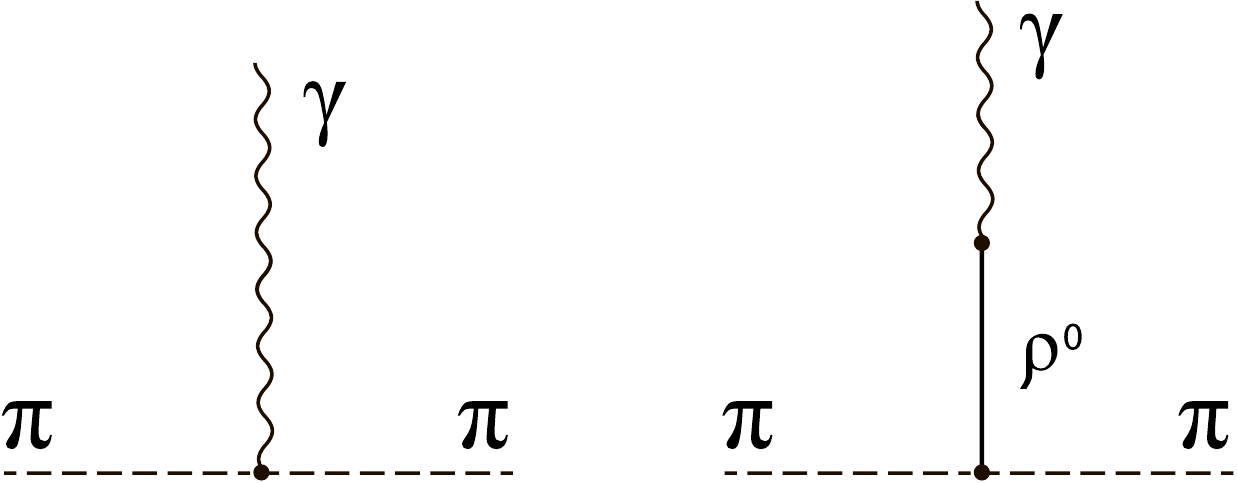}}
\caption{The Feynman diagrams describing the non-VMD and the VMD contributions to the $\gamma\pi\pi$ vertex in the covariant approach.}
\label{fig1}      
\end{figure}

Actually, $j_\mu^{em}$ contains an additional contribution to the $\gamma\pi\pi $ amplitude. It originates from the term
\begin{equation}
j_\mu^{em} \to -\frac{\kappa}{4G_V}\mbox{tr} \left(\xi_\mu Q\right) = -\frac{\kappa}{4G_V}  \xi_{\mu 3},
\end{equation}
giving
\begin{eqnarray}
\label{gpp1}
{\cal L}_{\gamma \pi\pi}^{(1)} &=&  -i \frac{e\kappa}{2G_V} {\cal A}_\mu \left(p^+\partial_\mu p^- - p^-\partial_\mu p^+\right) \nonumber \\
&=& -ie{\cal A}_\mu \left(\pi^+\partial_\mu \pi^- - \pi^-\partial_\mu \pi^+\right).
\end{eqnarray}
The first diagram in Fig.\ref{fig1} corresponds to this Lagrangian density. 

Let us obtain now the contribution which is generated by the VMD inspired mechanism applied to the $\rho\pi\pi$ vertex 
\begin{eqnarray}
{\cal L}_{\rho\pi\pi}&=& i\left( \frac{\kappa}{4G_V} -2g_AI_2\right) \mbox{tr} \left(v_\mu [p,\partial_\mu p ]\right) \nonumber \\
&-& i\, \frac{\kappa}{3}\, (1+g_A) I_2 \, \mbox{tr} \left( \tilde v_{\mu\nu} [\partial_\mu p, \partial_\nu p] \right).
\end{eqnarray}
Here, as we already know, the expression in the first parenthesis is zero. From the last term we extract the contribution of the neutral $\rho^0$-meson    
\begin{eqnarray}
{\cal L}_{\rho\pi\pi}=&-& i\,\frac{2}{3}g_\rho \kappa (1+g_A) I_2 \partial_\nu \tilde \rho^0_{\mu\nu} \nonumber \\
&\times& \left( p^+\partial_\mu p^- - p^-\partial_\mu p^+\right)+\ldots 
\end{eqnarray}
Now, the $\rho_\mu^0\to {\cal A}_\mu$ transition can be taken into account [see the second diagram in Fig.\ref{fig1}]. This leads to the simple replacements $g_\rho \to e, \rho_\mu^0\to {\cal A}_\mu $. Thus, it gives
\begin{eqnarray}
\label{gpp2}
{\cal L}_{\gamma\pi\pi}^{(2)}=&-&i \,\frac{2}{3}\, e\kappa (1+g_A) I_2 \partial_\nu {\cal F }_{\mu\nu}  
 \left( p^+\partial_\mu p^- - p^-\partial_\mu p^+ \right)   \nonumber \\
=&-&ie\,\frac{1+g_A}{2m^2_\rho} \partial_\nu {\cal F }_{\mu\nu} \left(\pi^+\partial_\mu \pi^- - \pi^-\partial_\mu \pi^+\right).
\end{eqnarray}
where ${\cal F}_{\mu\nu} = \partial_\mu {\cal A}_\nu -  \partial_\nu  {\cal A}_\mu $. This Lagrangian density does not contribute on the mass shell of the photon. 

Notice that the sum $\mathcal L_{\gamma\pi\pi}^{(1)}+\mathcal L_{\gamma\pi\pi}^{(2)}$ yields the same result as the standard approach, where only the second diagram contributes [this is because these two approaches lead to different $\rho\pi\pi$ vertices for off shell $\rho$-mesons]. It means that for this specific case there is no difference between standard and the gauge modified approach.

\subsection{$a_1 \pi\gamma$ amplitude}

\begin{figure}
\resizebox{0.20\textwidth}{!}{%
 \includegraphics{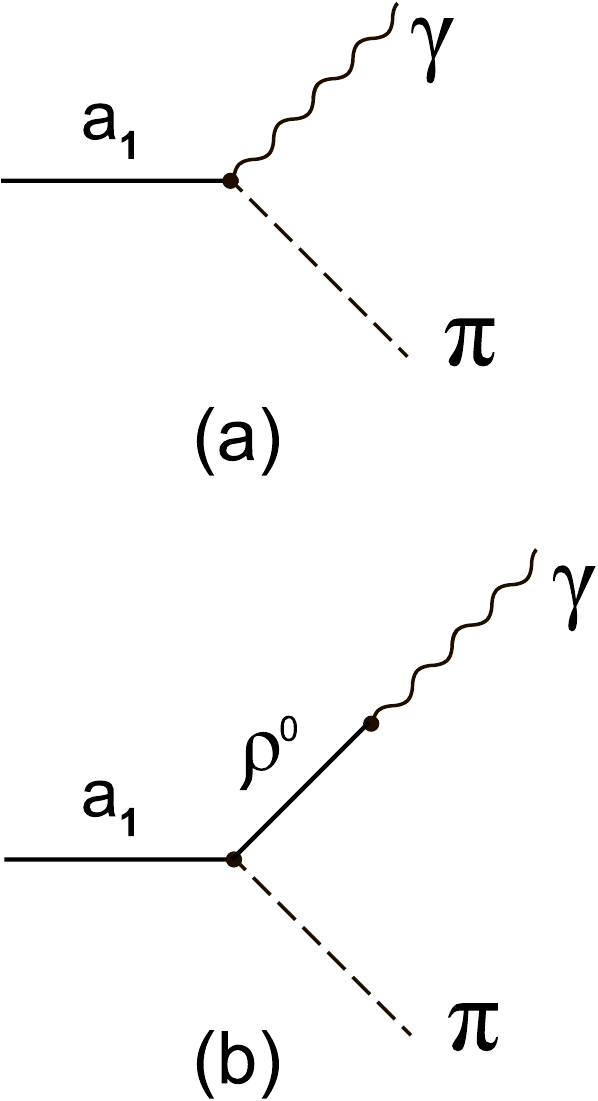}}
\caption{The Feynman diagrams describing the vertex $a_{1}(1260) \to \gamma\pi $ in the covariant approach.}
\label{fig2}      
\end{figure}

We now proceed to calculate the $a_1\pi\gamma$ coupling which is described by the diagrams shown in Fig.\ref{fig2}. To find the contribution of the first diagram, we should consider the current $j_\mu^{em}$. The relevant terms in $j_\mu^{em}$ are
\begin{eqnarray}
j_\mu^{em}&\to& i\kappa \left( \frac{1}{G_V} +16m^2 I_2\right) \mbox{tr}\left(\Psi^{em} a_\mu \right)\nonumber\\
&-&4i\kappa \frac{2}{3}I_2 \mbox{tr} \left(\partial_\nu \tilde a_{\mu\nu} \Psi^{em}\right) \nonumber \\
&=&\frac{4i\kappa}{g_\rho^2}\mbox{tr}\left[\left(m^2_{a_1} a_\mu - \partial_\nu \tilde a_{\mu\nu} \right)\Psi^{em}\right],
\end{eqnarray}
where $\tilde a_{\mu\nu}=\partial_\mu a_\nu - \partial_\nu a_\mu $, and the following model relations have been used
\begin{equation}
\frac{1}{G_V} +16m^2 I_2 = \frac{4m^2_{a_1}}{g_\rho^2}, \quad   \frac{2}{3}I_2=\frac{1}{g_\rho^2}.
\end{equation}
Observing, that
\begin{equation}
\mbox{tr}\left(a_\mu \Psi^{em}\right) \to -\frac{m}{2}\mbox{tr} \left( a_\mu [Q,p] \right),
\end{equation}
we can write down the corresponding Lagrangian density
\begin{equation}
\label{apg1}
{\cal L}^{(1)}_{a_1\pi\gamma}=-6ie\frac{m}{g_\rho^2}\mbox{tr} \left\{ \left( a_\mu -  \frac{\partial_\nu \tilde a_{\mu\nu}}{m^2_{a_1}}  \right)  [\bar {\cal A}_\mu, p]  \right\},
\end{equation}
where $\bar {\cal A}_\mu = {\cal A}_\mu Q$ and we have used the formula $\kappa =3/ m^2_{a_1}$, which follows from (\ref{pac}) and mass equations (\ref{mrho}) - (\ref{mrel}).

One can obtain the VMD contribution to this process [see Fig.\ref{fig2}b] by considering the Lagrangian density for the vertex $a_1\rho^0\pi$ in ${\cal L}_{SD}$, and taking into account the $\rho^0_\mu \to {\cal A}_\mu$ transition described by Eq.(\ref{vmd}). We leave out some of the algebra and just write the resulting expression
\begin{eqnarray}
\label{apg2}
&& {\cal L}_{SD}\to {\cal L}^{(2)}_{a_1\pi\gamma} = 6ie \frac{m}{g_\rho^2}\mbox{tr}\left\{ a_\mu [\bar {\cal A}_\mu, p] \right. \nonumber \\
&+&\left. \frac{1}{m^2_{a_1}} \left(\tilde a_{\mu\nu} [\bar {\cal A}_\mu, \partial_\nu p]
-a_\mu [\bar {\cal F}_{\mu\nu}, \partial_\nu p ] \right)\right\}.
\end{eqnarray}
Here, $\bar {\cal F}_{\mu\nu} = \partial_\mu \bar {\cal A}_\nu -  \partial_\nu \bar {\cal A}_\mu $. Upon integrating by parts in the corresponding action we find that the sum of the two contributions (\ref{apg1}) and (\ref{apg2}) can be expressed as
\begin{eqnarray}
\label{a1pg}
{\cal L}_{a_1\pi\gamma} &=& \frac{6iem}{g_\rho^2 m^2_{a_1}} \mbox{tr} \left(a_\mu [ \partial_\nu \bar {\cal F}_{\mu\nu}, p] \right) \nonumber \\
&=& \frac{ieg_\rho f_\pi }{2m_\rho^2} \mbox{tr} \left(a_{1\mu} [ \partial_\nu \bar {\cal F}_{\mu\nu}, \pi] \right),
\end{eqnarray}
where on the last step we introduced the physical fields in accord with Eqs. (\ref{Rsp}) and (\ref{Rvach}); $a_{1\mu}\to\vec a_{1\mu} \vec \tau$.

Since the Lagrangian density (\ref{a1pg}) vanishes on the $\gamma$-mass-shell, we conclude that the $a_1 \pi\gamma$ vertex may contribute for the virtual photons. As opposed to this, that vertex is given in the non-covariant scheme by the expression (\ref{apg2}), which vanishes on the $a_1$-meson mass shell. So, in this case, we obtain a different off-shell behaviour, however, both results are zero on the mass shell of the photon and the $a_1$-meson.

\subsection{The $\rho^0\to \gamma\pi^+\pi^-$ amplitude}

Let us consider now a more sophisticated  example. This is the $\rho^0(p)\to \gamma (q)+\pi^+ (p_+)+\pi^-(p_-)$ decay, where $p, q, p_+, p_-$ are the 4-momenta of the corresponding particles. Our goal is to show that the resulting amplitude does not depend on whether one uses the covariant or the standard approach.
 
In general, the decay amplitude obtains two types of contributions. These are the emission of the photon by a single charged pion, and gamma emission from the core of the hadronic region. 

In both models, the first process is described by the same vertices, giving the following result
\begin{equation}
A_{\pi}=eg_\rho (1+g_A) \epsilon_\mu (p)\epsilon^*_\nu (q)\left(\frac{p_+^\mu p_-^\nu}{qp_-} +\frac{p_-^\mu p_+^\nu}{qp_+}\right), 
\end{equation} 
where $\epsilon_\mu (p)$ and $\epsilon_\nu (q)$ are polarization vectors of the $\rho^0$-meson and the photon, correspondingly.   

\begin{figure}
\resizebox{0.25\textwidth}{!}{%
 \includegraphics{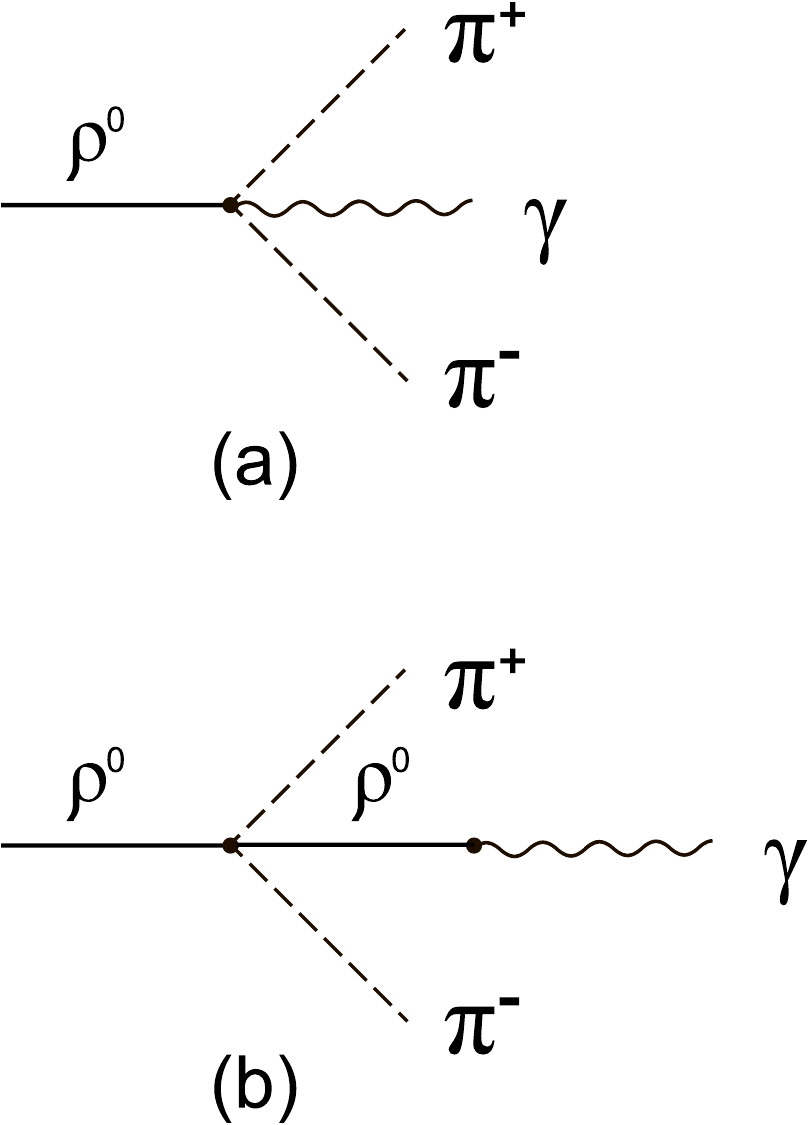}}
\caption{The Feynman diagrams describing the $\rho^{0} \to \gamma \pi^{+} \pi^{-}$ amplitude in the covariant approach.}
\label{fig3}      
\end{figure}

Let us now turn to the emission of the photon from the internal region. In the covariant approach, this process is described by the diagrams shown in Fig.\ref{fig3}. Notice the lack of the $a_1$-exchange: as we already know from (\ref{a1pg}) the $a_1\to\pi\gamma$ amplitude vanishes on the mass shell of the photon. The appropriate Lagrangian densities, in general, contain the terms without derivatives and with two derivatives. 

The constant term of the first diagram (Fig.\ref{fig3}a) cancels the constant term of the second one (Fig.\ref{fig3}b). Indeed, the first diagram represents the contribution of the electromagnetic current $j_\mu^{em}$
\begin{eqnarray}
\label{vppg}
j_\mu^{em}&\to &\frac{1}{2G_V}\mbox{tr}\left(2\kappa v_\mu\Xi^{em} \right) \nonumber \\
   &-& 8\kappa m I_2 \mbox{tr}\left(\Psi^{em} [v_\mu , p] \right)  \nonumber \\
   &\to& -2\kappa \left(\frac{1}{G_V}+16m^2 I_2 \right) v_{\mu 3} p^+p^- \nonumber \\
   &=& -16I_2 v_{\mu 3} p^+p^-
\end{eqnarray} 
The contribution without derivatives, described by the second diagram, originates only from one term of the Lagrangian density (\ref{LSD}), namely
\begin{equation}
\bigtriangledown_\mu p \to  -i[v_\mu , p].
\end{equation}
The corresponding part of the Lagrangian density is given by
\begin{equation}
\label{ccca}
{\cal L} \to - I_2 \mbox{tr}  [v_\mu , p]^2  =16I_2 v_{\mu 3}^2p^+p^-.
\end{equation}
Turning on the $\rho^0_\mu\to {\cal A}_\mu$ VMD transition in (\ref{ccca}) one can easily see that this term cancels the contact contribution of the electromagnetic current (\ref{vppg}). 

The term with two derivatives from $j_\mu^{em}$, corresponding to the diagram of Fig.\ref{fig3}a, is given by    
\begin{eqnarray}
\mathcal L^{(a)}&=&\frac{16}{3} e\kappa I_2 \mathcal A_\mu \left\{ \partial_\nu \tilde v_{\mu\nu 3} p^+ p^- +\kappa m^2 [\tilde v_{\mu\nu 3} \partial_\nu (p^+ p^-)\right. \nonumber \\
&-&\partial_\nu v_{\mu 3} \partial_\nu (p^+ p^-)- v_{\mu 3}\left(p^+\partial^2_\nu p^- + p^-\partial^2_\nu p^+   \right) \nonumber \\ 
&+&  v_{\nu 3}\left(p^+\partial_\nu \partial_\mu p^- + p^-\partial_\nu \partial_\mu p^+ \right) \nonumber \\
&+&\left. \partial_\nu v_{\nu 3} \partial_\mu (p^+ p^-)] \right\},
\end{eqnarray}  
where $\tilde v_{\mu\nu 3}=\partial_\mu v_{\nu 3}-\partial_\nu v_{\mu 3}$ is the strength tensor. 

The two derivatives part of the second diagram (Fig. \ref{fig3}b) is described by the following Lagrangian density
\begin{eqnarray}
\mathcal L^{(b)} &=& \frac{16}{3}e\kappa^2 m^2 I_2 \mathcal A_\mu\left[ -2v_{\mu 3}\partial_\nu p^+ \partial_\nu p^- \right.  \nonumber \\
&+&\left. v_{\nu 3} \left(\partial_\mu p^+ \partial_\nu p^-  + \partial_\nu p^+ \partial_\mu p^- \right)\right].
\end{eqnarray}

After some rearrangement of derivatives and omitting total derivatives, we obtain that the sum of these two Lagrangians on the mass shell of the photon is  
\begin{equation}
\mathcal L^{(a)}+\mathcal L^{(b)}=\frac{8}{3} e\kappa (1+g_A) I_2 \mathcal A_\mu \partial_\nu \tilde v_{\mu\nu 3} p^+p^-
\end{equation}
On the mass shell of the $\rho^0$-meson $\partial_\nu \tilde v_{\mu\nu 3} =m_\rho^2 v_{\mu 3}$, and we finally find for the sum $\mathcal L^{(a+b)}=\mathcal L^{(a)}+\mathcal L^{(b)}$ a very simple expression 
\begin{eqnarray}
\label{a+b}
\mathcal L^{(a+b)}&=& 8e I_2 g_A (1+g_A) \mathcal A_\mu v_{\mu 3} p^+p^- \nonumber \\
&=& eg_\rho  (1+g_A) \mathcal A_\mu \rho^0_{\mu} \pi^+\pi^-.
\end{eqnarray}

This Lagrangian density leads to an additional contact contribution to the total amplitude of the $\rho^0 \to \gamma\pi^+\pi^-$ decay. As a result, we have 
\begin{equation}
\label{A}
A =eg_\rho (1+g_A) \epsilon_\mu (p)\epsilon^*_\nu (q)\left( g^{\mu\nu}+ \frac{p_+^\mu p_-^\nu}{qp_-} +\frac{p_-^\mu p_+^\nu}{qp_+}\right). 
\end{equation} 
One can easily see that this amplitude is gauge invariant. Indeed, replacing $\epsilon^*_\nu (q)\to q_\nu$ one finds that    
\begin{equation}
\epsilon_\mu (p)q_\nu\left( g^{\mu\nu}+ \frac{p_+^\mu p_-^\nu}{qp_-} +\frac{p_-^\mu p_+^\nu}{qp_+}\right)=\epsilon_\mu (p) p^\mu=0
\end{equation} 
vanishes on the $\rho^0$ mass shell.  

\begin{figure}
\resizebox{0.35\textwidth}{!}{%
 \includegraphics{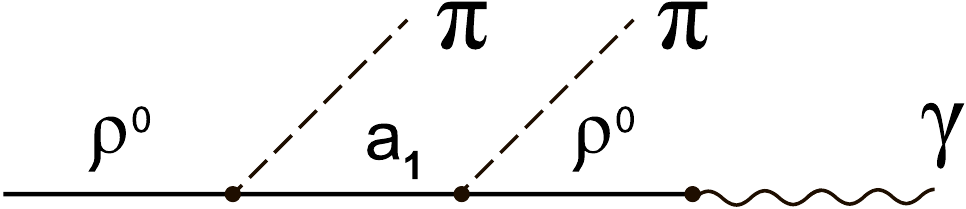}}
\caption{The additional Feynman diagrams describing the VMD contribution to the $\rho^{0}\to \gamma \pi^{+} \pi^{-}$ amplitude in the non-covariant approach.}
\label{fig4}      
\end{figure}

Notice that the diagram Fig.\ref{fig3}a in the covariant approach plays the role of the diagram Fig.\ref{fig4} in the standard case. Indeed, the standard approach gives the same Lagrangian for the diagram Fig.\ref{fig3}b, but does not have the diagram Fig.\ref{fig3}a. If we calculate the contribution of the $a_1$ exchange diagram (Fig.\ref{fig4}), we find    
\begin{eqnarray}
A_{a_1}&=&eg_\rho \frac{1-Z}{m^2_{a_1}} \epsilon_\mu (p)\epsilon^*_\nu (q) \left\{  \right.\nonumber \\
&& [m^2_{a_1} +p^2 -(q+p_+)^2]g^{\mu\nu} -p_-^\mu p_+^\nu \nonumber\\
&+&\left.  [m^2_{a_1} +p^2 -(q+p_-)^2]g^{\mu\nu} -p_+^\mu p_-^\nu \right\}.
 \end{eqnarray}
Combining this result with the amplitude given by the diagram Fig.\ref{fig3}b, we obtain the known result (\ref{a+b}), or after including the gamma emissions from the charged pions the result (\ref{A}). 

This example helps to understand the role of non-VMD contributions in the covariant approach: they effectively reproduce the result of the VMD approach. One can show \cite{OHZ18} that this always take place if we are working with the real part of effective action. However, one cannot extend this statement to the imaginary part of the action.       

\section{One $Z$-boson interactions}
\label{Zbi}

Our next goal concerns the hadronic interactions of the neutral $Z$-boson. The leading order [in coupling $g$] interactions can be presented by the Lagrangian density
\begin{equation}
\label{LZ}
{\cal L}_Z =g Z_\mu j^{N}_\mu,
\end{equation}
where $j^{N}_\mu$ is the corresponding hadron neutral current.

It is useful to introduce again two matrices which help us to write the current $j^{N}_\mu$. One of them is the matrix $\Xi^N$, which is defined as
\begin{equation}
\Xi \to \frac{2gZ_\mu}{\cos\theta_W}\Xi^N,
\end{equation}
where
\begin{eqnarray}
\Xi^N&=&\cos 2\theta_W \Xi^{em} \nonumber \\
         &+&\frac{i}{\sqrt{2}} \left( \begin{array}{cc} 0 &  p^+\bar s_0-s^+p_0 \\
                p_0s^- -p^-\bar s_0  &   0  \end{array}\right) .
\end{eqnarray}

The other matrix is $\Psi^N$. It collects the hadron part of $\Psi_\mu$ which is linear in $Z$.
\begin{equation}
\Psi_\mu \to \frac{2gZ_\mu}{\cos\theta_W} \Psi^N,
\end{equation}
with
\begin{eqnarray}
&&\Psi^N=\cos 2\theta_W\,  i\Psi^{em} \nonumber \\
&&+\left( \begin{array}{cc} p_0p_3+\frac{1}{2}(p_0^2+p_3^2) &  \frac{1}{\sqrt{2}}p^+p_3 \\
               \frac{1}{\sqrt{2}}p^-p_3  &   p_0p_3-\frac{1}{2}(p_0^2+p_3^2)  \end{array}\right)  \nonumber\\
&& +\left( \begin{array}{cc} \bar s_0s_3+\frac{1}{2}(\bar s_0^2+s_3^2) &  \frac{1}{\sqrt{2}}s^+s_3 \\
               \frac{1}{\sqrt{2}}s^-s_3  &  \bar s_0s_3-\frac{1}{2}(\bar s_0^2+s_3^2)  \end{array}\right),
\end{eqnarray}
and
\begin{equation}
\bar\Psi^N =\Psi^N-m^2 T_3.
\end{equation}
The latter formula takes into account the subtraction of $Z_\mu$ term made in Eq.(\ref{psibar}).

These notations allow us to express the current $j^{N}_\mu$ as follows
\begin{eqnarray}
\label{Ncur}
j_\mu^{N}&=& \frac{1}{4G_V\cos\theta_W}\left\{
2\kappa \mbox{tr} \left[\bar \Psi^{N} \left(a_\mu+\frac{\kappa}{2}\zeta_\mu \right)\right.\right. \nonumber \\
&+&\left. \Xi^{N}\left(v_\mu +\frac{\kappa}{2} \xi_\mu \right)\right] +\frac{v_{\mu 0}}{3} \sin^2\theta_W    \nonumber \\
&-&\left.\left(v_{\mu 3}+\frac{\kappa}{2}\xi_{\mu 3} \right) \cos^2\theta_W + a_{\mu 3} +\frac{\kappa}{2}\zeta_{\mu 3}\right\}     \nonumber\\
&+&\frac{2\kappa I_2}{\cos\theta_W}  \mbox{tr}\left\{ \overline{\bigtriangledown_\mu p} \left( \{\bar\Psi^{N}, \bar s\}-i[\Xi^{N},p]  \right)\right. \nonumber \\
&-& \overline{\bigtriangledown_\mu s} \left( \{\bar\Psi^{N}, p\}+i[\Xi^{N},s]  \right) + i\,\frac{2}{3}\,\overline{v}_{\mu\nu}\nonumber \\
&\times&\left( [ \Xi^{N},v_\nu + \frac{\kappa}{2}\xi_\nu ] + [\bar\Psi^{N}, a_\nu +\frac{\kappa}{2}\zeta_\nu ] \right)
      + i\,\frac{2}{3}\,\overline{a}_{\mu\nu} \nonumber\\
&\times& \left( [\Xi^{N}, a_\nu +\frac{\kappa}{2}\zeta_\nu ]  + [\bar\Psi^{N}, v_\nu + \frac{\kappa}{2}\xi_\nu ] \right)\nonumber\\
&-&\left. \frac{2}{3}\left( \partial_\nu \overline{v}_{\mu\nu} \Xi^{N}+ \partial_\nu \overline{a}_{\mu\nu} \bar\Psi^{N}  \right)\right\}.
\end{eqnarray}
We emphasize that currents (\ref{jem}) and (\ref{Ncur}) are written in terms of fields which still must be redefined in accord with Eqs. (\ref{Rsp})-(\ref{Rvach}).

\subsection{Neutral current-current approximation}

At low momenta of the $Z$-boson, $p^2\ll m_Z^2$, one can replace the $Z$-propagator by a constant value multiplied by the delta function and the metric tensor
\begin{equation}
\Delta_{\mu\nu}^{(Z)} (x_1-x_2)\to i\,\frac{g_{\mu\nu}}{m_Z^2}\delta (x_1-x_2)
\end{equation}
In this approximation, we can derive the local current-current interaction between the leptonic weak neutral current $l_\mu^N$ and hadronic neutral current $j_\mu^N$. Indeed, in the Weinberg-Salam model the coupling of the $Z$-boson with $l^N_\mu$ can be written as
\begin{equation}
{\cal L}_{ZN}=\frac{g}{4\cos\theta_W}Z_\mu l_\mu^N,
\end{equation}
where
\begin{equation}
l_\mu^N =\!\!\! \sum_{\ell =e,\mu,\tau} \left[ \bar \nu_{\ell} \gamma_\mu (1-\gamma_5) \nu_{\ell} +\bar \ell \gamma_\mu \left(4\sin^2\theta_W -1+\gamma_5 \right) \ell   \right]
\end{equation}
Combining this result with Eq. (\ref{LZ}), we obtain
\begin{equation}
{\cal L}_{NN }=-\frac{G_F}{\sqrt{2}}\, 2\cos\theta_W\, l_\mu^N j_\mu^N.
\end{equation}
If one takes only the one-particle part in $j_\mu^N$, it gives
\begin{eqnarray}
\label{lnn}
{\cal L}_{NN}=&-&\frac{G_F}{\sqrt{2}}\left[\frac{m_\rho^2}{g_\rho}\left(\frac{\omega_\mu}{3}\sin^2\theta_W - \rho_\mu^0 \cos^2\theta_W +a^0_\mu \right) \right. \nonumber \\
&+&\left. f_\pi \partial_\mu \pi^0 \right] l^N_\mu + \ldots
\end{eqnarray}
This expression contains the parity-violating $\Delta I=1$ changing isospin interactions. For instance, the last term in (\ref{lnn}) yields the $e^+e^-\pi^0$-vertex with the factor $4\sin^2\theta_W -1$. Notice that this small factor would vanish for $\sin^2\theta_W = 0.25$.

\subsection{Effect of P violation}

One of the interesting consequences of the current (\ref{Ncur}) is that it generates the flavor-conserving $\Delta I=1$  parity-violating transitions in hadron processes. In particular, the effect of P violation in the $\Delta S = 0$,  $\Delta I=1$ channel is carried by the part of the Lagrangian density (\ref{zd}) which is responsible for the weak $\omega\to a_{1}^0$ transition. The amplitude for this process is
\begin{equation}
A=\left(\frac{igm_\rho^2}{2\cos\theta_W g_\rho}\right)^2\frac{1}{3}\sin^2\theta_W \epsilon_\mu(p)
\epsilon^*_\nu (p)\frac{g_{\mu\nu}-\frac{p_\mu p_\nu}{m_Z^2}}{m_Z^2-p^2}
\end{equation}

For $p^2 \ll m_Z^2$  we can replace the $Z$-propagator by the constant value $g_{\mu\nu}/m_Z^2$, leading to the amplitude
\begin{eqnarray}
A&\to& \left(\frac{igm_\rho^2}{2m_Z\cos\theta_W g_\rho}\right)^2\frac{1}{3}\sin^2\theta_W \epsilon_\mu(p)
\epsilon^*_\mu (p) \nonumber \\
&=& -\frac{G_F}{\sqrt{2}}\left(\frac{m_\rho^2}{g_\rho} \right)^2 \frac{2}{3} \sin^2\theta_W \epsilon_\mu(p)
\epsilon^*_\mu (p),
\end{eqnarray}
where $G_F$ is the Fermi coupling constant, which is given by $G_F/\sqrt{2}=g^2/(8m_W^2)$. This corresponds to the Lagrangian density
\begin{equation}
{\cal L}_{\omega a_1}=-\frac{G_F}{\sqrt{2}}\left(\frac{m_\rho^2}{g_\rho} \right)^2 \frac{2}{3} \sin^2\theta_W \omega_\mu (x) a_{1\mu}^0 (x).
\end{equation}

\section{One $W$-boson interactions}
\label{Wbi}

We will discuss now the structure of the charged hadronic current $j_\mu^C$, which is responsible for the single $W$-boson interactions of mesons. In this case, the Lagrangian density can be presented in the form
\begin{equation}
\label{LW}
{\cal L}_{W}=g\left(W_\mu^+ j_\mu^C+ \mbox{h.c.}\right).
\end{equation}
Similar to the previous cases, the charged hadronic current can be expressed in terms of two meson matrices $\Xi^C$ and $\bar \Psi^C=\Psi^C -2m^2T_+$. These matrices accumulate the meson part of $\Xi_\mu$ and $\Psi_\mu$ which is responsible for the non-axial-vector dominant interactions of mesons with $W^\pm$. They are extracted as follows
\begin{equation}
   \Xi_\mu \to g W^+_\mu \Xi^C, \quad  \Psi_\mu \to g W^+_\mu \Psi^C.
\end{equation}
In accord with this prescription, we find that the matrix elements of $\Xi^C$ are
\begin{eqnarray}
\Xi^C_{11}&=& p^-(p_3+i\bar s_0)+s^-(s_3-ip_0),   \nonumber\\
\Xi^C_{12}&=&-\sqrt{2}\left[p_3(p_3+i\bar s_0)+s_3 (s_3-ip_0) +p^+p^-+s^+s^- \right], \nonumber \\
\Xi^C_{21}&=&   \sqrt{2}(p^+p^-+s^+s^-),            \nonumber\\
\Xi^C_{22}&=& - \Xi^C_{11}.
\end{eqnarray}
The matrix $\Psi^C$ has the following elements
\begin{eqnarray}
\Psi^C_{11}&=& 2\left[p_0p^- +\bar s_0s^- +i(p^-s_3-s^-p_3)\right] \nonumber \\
&+& p_3p^-+s_3s^-+i(\bar s_0p^--p_0s^-),  \nonumber\\
\Psi^C_{12}&=& \sqrt{2}\left[p^+p^-+ p_0^2+\bar s_0^2+s^+s^- +i( p_0s_3-\bar s_0p_3  )\right],   \nonumber\\
\Psi^C_{21}&=& \sqrt{2}(p^-p^-+s^-s^-),   \nonumber\\
\Psi^C_{22}&=& 2\left[p_0p^- +\bar s_0s^- +i(p^-s_3-s^-p_3)\right] \nonumber \\
&-& p_3p^--s_3s^--i(\bar s_0p^--p_0s^-).
\end{eqnarray}

Finally, let us recall the formula (\ref{Ltot}). Collecting the appropriate terms from ${\cal L}_{tot}$, and using our definitions above, we can write down the expression for the current $j_\mu^C$
\begin{eqnarray}
\label{Ccur}
j_\mu^{C}
&=& \frac{1}{4G_V}\left\{ a^-_{\mu} - v^-_{\mu} +\frac{\kappa}{2}\left( \zeta_{\mu -} - \xi_{\mu -}\right) \right. \nonumber \\
&+&\left. \kappa\, \mbox{tr} \left[ \bar \Psi^{C} \left(a_\mu+\frac{\kappa}{2}\zeta_\mu \right) +\Xi^{C}\left(v_\mu +\frac{\kappa}{2} \xi_\mu \right)\right]   \right\}      \nonumber \\
&+&\kappa I_2\, \mbox{tr}\left\{ \overline{\bigtriangledown_\mu p} \left( \{\bar\Psi^{C}, \bar s\}-i[\Xi^{C},p]  \right)\right. \nonumber \\
&-& \overline{\bigtriangledown_\mu s} \left( \{\bar\Psi^{C}, p\}+i[\Xi^{C},s]  \right) + i\,\frac{2}{3}\,\overline{v}_{\mu\nu}   \nonumber \\
&\times&\left( [ \Xi^{C},v_\nu + \frac{\kappa}{2}\xi_\nu ] + [\bar\Psi^{C}, a_\nu +\frac{\kappa}{2}\zeta_\nu ] \right)
      + i\,\frac{2}{3}\,\overline{a}_{\mu\nu} \nonumber\\
&\times& \left( [\Xi^{C}, a_\nu +\frac{\kappa}{2}\zeta_\nu ]  + [\bar\Psi^{C}, v_\nu + \frac{\kappa}{2}\xi_\nu ] \right)\nonumber\\
&-&\left. \frac{2}{3}\left( \partial_\nu \overline{v}_{\mu\nu} \Xi^{C}+ \partial_\nu \overline{a}_{\mu\nu} \bar\Psi^{C}  \right)\right\}.
\end{eqnarray}
This current, for instance, can be readily used for the calculation of the various decay rates of the $\tau$ meson. For that the following current-current approximation is useful.

\subsection{Charged current-current approximation}

Let us consider now the charged leptonic current $l_\mu^C$ and its Hermitian conjugate one $l_\mu^{C\dagger}$
\begin{equation}
l_\mu^C= \bar\ell \gamma^\mu (1-\gamma_5)\nu_\ell , \quad
l_\mu^{C\dagger}= \bar\nu_\ell \gamma^\mu (1-\gamma_5)\ell .
\end{equation}
These phenomenological currents of the earlier current-current Fermi model are precisely those used in the gauge $SU(2)_L$ theory. Indeed, the charge-changing part of the standard model Lagrangian density has the form
\begin{equation}
\label{WC}
{\cal L}_{WC}=\frac{g}{2\sqrt{2}}\sum_{\ell =e,\mu,\tau} W_\mu^- l_\mu^C  + \mbox{h.c.}
\end{equation}

With the help of Eqs. (\ref{LW}) and (\ref{WC}) we can return back to the old current-current type Lagrangian density. The method used remains the same as the one used in the previous section. As a result we find
\begin{equation}
{\cal L}_{CC}=-2G_F \sum_{\ell =e,\mu,\tau} l_\mu^C  j_\mu^C + \mbox{h.c.}
\end{equation}
Correspondingly, the one-particle part of the charged weak hadron current $j_\mu^C$ has the following contribution to the effective Lagrangian density
\begin{eqnarray}
{\cal L}_{CC}=&-&G_F\!\!\!\sum_{\ell =e,\mu,\tau}  l_\mu^C  \left[ \frac{m_\rho^2}{g_\rho} \left(a_{1\mu}^- -\rho_\mu^- \right) +f_\pi \partial_\mu \pi^- +\ldots  \right] \nonumber\\
&+&  \mbox{h.c.}
\end{eqnarray}
Notice, that the current $j_\mu^C$ contains also many-particle direct contributions [indicated as a plus followed by suspension points] which are important for the multi-meson production reactions.

\section{Conclusions}
\label{concl}

Starting from an extended $U(2)_L\times U(2)_R$ chiral symmetric NJL model with explicit flavour symmetry breaking we have derived an effective meson Lagrangian which describes both the strong and electroweak low-energy interactions of hadrons. This theory describes the phenomenon of dynamical chiral symmetry breaking. As a result, a non-zero quark anti-quark condensate is generated and the strong dynamics of the system is described by small bosonic excitations (mesons) in the Nambu-Goldstone ground state. This phenomenon has many interesting consequences widely discussed in the literature. In our study we concentrated on the consistent treatment of the $\pi -a_{1}$ mixing. We have shown that in presence of the electroweak interactions the elimination of mixing terms should be performed in accord with the general requirements of chiral and electroweak symmetries. Somehow this point was totally ignored in the literature. We suggested a self-consistent procedure of covariant diagonalization of the mixed terms and obtained the corresponding effective meson Lagrangian. This Lagrangian contains the essential features of the standard meson dynamics [Goldberger-Treiman relation, KSFR relation, approximate Weinberg relation, PCAC, field-current identities for electro-weak current, universality of induced meson coupling constants] and gives rise to some additional consequences. One of them is a deviation from the vector and axial-vector meson dominance picture. This deviation is a direct consequence of the covariant approach, which has been the main focus of the present work. 

One of the interesting future applications of the obtained theory is the study of its anomaly sector. Our previous investigations have shown that new non-VMD contributions are responsible for the restoration of the gauge symmetry for the $a_1\to\gamma\pi^+\pi^-$ decay amplitude. However, it would be also important to clarify, in the study of anomalies, the role of the new contributions, which originate in the weak sector and induce deviations of the theory from the axial-vector dominance of the charged hadronic current.

Furthermore, one should mention that the extension of the chiral group to the $SU(3)_L\times SU(3)_R$ case will allow us to apply the idea of gauge invariant $p-a_\mu$ diagonalization to the strange quark physics.

\section*{Acknowledgments}
A.A.O. is grateful for warm hospitality at the IMP of the Chinese Academy of Sciences in Lanzhou. P.M.Z is supported by the National Natural Science Foundation of China (Grant No. 11575254). B.H. acknowledges CFisUC and FCT through the project UID/FIS/04564/2016. We acknowledge networking support by the COST Action CA16201.

\end{document}